\documentclass[graybox, vecphys, natbib]{svmult}

\usepackage{mathptmx}       
\usepackage{helvet}         
\usepackage{courier}        
\usepackage{type1cm}        
\usepackage{makeidx}         
\usepackage{graphicx}        
\usepackage{multicol}        
\usepackage[bottom]{footmisc}
\usepackage{hyperref}        
\usepackage{soul}            

\makeindex             
                       
\begin{document}
\bibliographystyle{spbasic}
\title*{Nucleosynthesis in neutrino-heated ejecta and neutrino-driven winds of core-collapse
supernovae; neutrino-induced nucleosynthesis}
\titlerunning{Nucleosynthesis in neutrino-driven ejecta of core-collapse supernovae}
\author{Shinya Wanajo}
\institute{First Author \at Max-Planck-Institut f\"ur Gravitationsphysik (Albert-Einstein-Institut), Am M\"uhlenberg 1, D-14476 Potsdam-Golm, Germany, \email{shinya.wanajo@aei.mpg.de}}
%
%
\maketitle
\abstract{The innermost ejecta of core-collapse supernovae are considered to be the sources of some iron-group and trans-iron nuclei. The ejecta are predominantly driven by neutrino heating, principally due to neutrino capture on free neutrons and protons. Such neutrino interactions play a crucial role for setting neutron richness in the ejecta. Recent hydrodynamical work indicates that the ejecta are only mildly neutron rich, or even proton rich. Under such conditions a wide variety of iron-group and trans-iron isotopes are synthesized, while the inferred neutron richness appears insufficient for producing r-process nuclei. In this chapter, basic concepts of nucleosynthesis in neutrino-heated ejecta and 
neutrino-driven winds of core-collapse supernovae are presented along with some latest results in the literature. Here, ``neutrino-heated ejecta" are referred to the early component within the first few seconds, in which anisotropic convective activities of material above the neutron star surface become important for nucleosynthesis. These are followed by ``neutrino-driven winds", which are approximately isotropic outflows emerging from the surface of a proto-neutron star. For these reasons, studies of nucleosynthesis presented here are based on recent multi-dimensional hydrodynamical simulations and semi-analytic wind solutions, respectively. These studies suggest that light trans-iron species up to atomic mass number of 90 are produced in neutrino-heated ejecta. Neutrino-driven winds are unlikely sources of r-process nuclei, but rather promising sources of proton-rich isotopes up to atomic number of 110.}

\section{\textit{Introduction}}
\label{sec:introduction}

\begin{figure}
    \centering
    \includegraphics[width=0.8\textwidth]{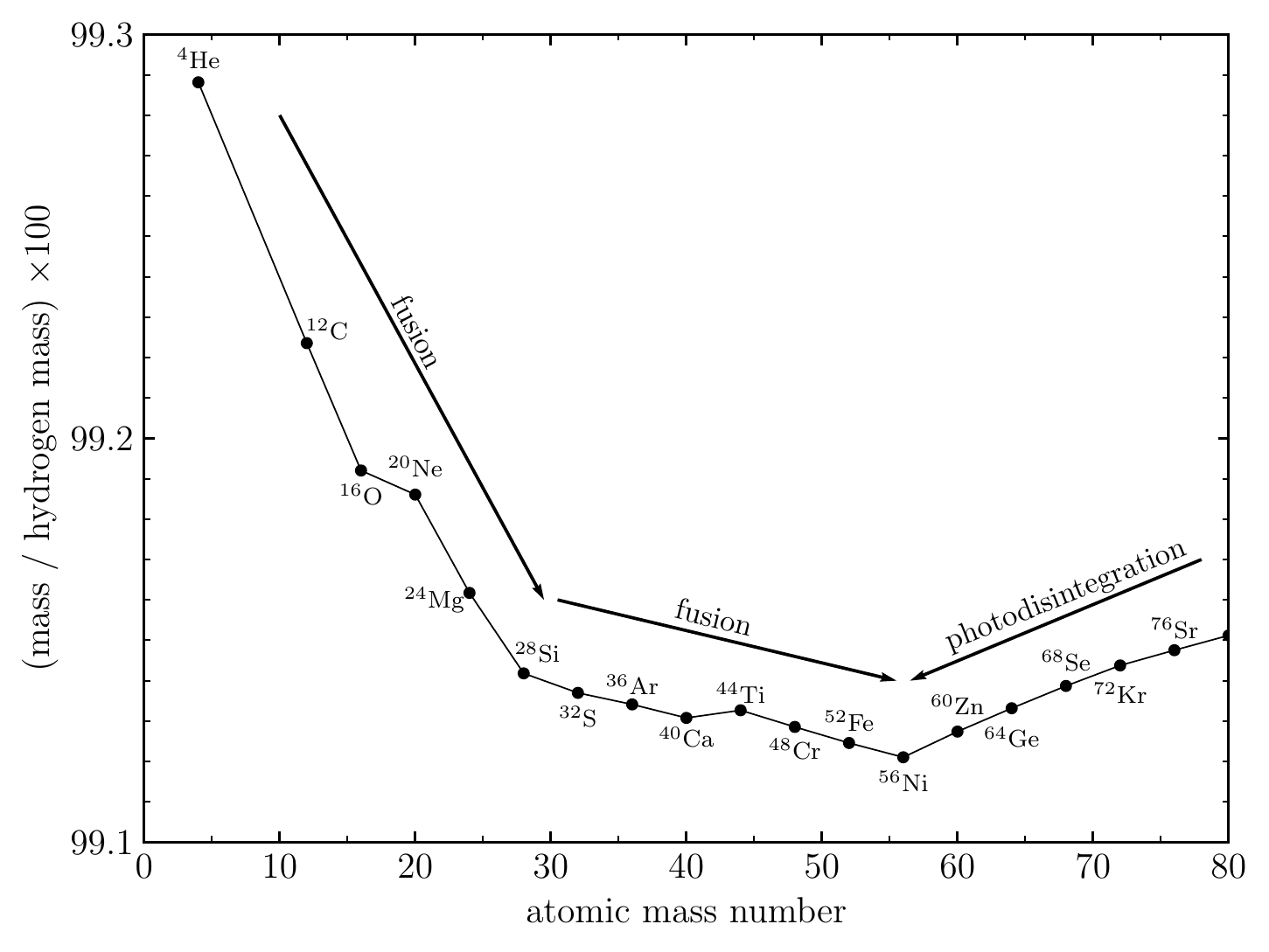}
    \caption{Valley of stability for $\alpha$ nuclei (consisting of multiple $^4$He). Nuclear masses ($\times 100$) with respect to that of hydrogen are plotted as a function of atomic number. In the central core of a massive star at the final stage, the end product by nuclear fusion is $^{56}$Ni because it has the smallest mass.}
    \label{fig:mass_alpha}
\end{figure}

Stars with masses at birth above $\sim 8\, M_\odot$, where $M_\odot$ indicates the solar mass, end their lives as energetic explosions known as core-collapse supernovae \citep{Janka2012}. Just prior to an explosion, the central core with a radius of $\sim 1000$~km (or with an enclosed mass of $\sim 1.4\, M_\odot$), consisting mainly of $^{56}$Ni and some other iron-group species, is supported by pressure stemming from the high temperature ($\sim 5$~GK, where GK $= 10^9$~K) and density ($\sim 10^9$~g~cm$^{-3}$). However, further gravitational contraction of the core cannot generate nuclear energy owing to the smallest mass of $^{56}$Ni among $\alpha$ nuclei (made of multiple $^4$He), the main constituents of a stellar core (Fig.~\ref{fig:mass_alpha}). This results in photodisintegration of $^{56}$Ni into $\alpha$ particles and free nucleons (neutrons and protons) as well as electron capture on nuclei. These photodisintegration and electron capture processes abruptly remove the pressure support at the center, leading to gravitational collapse of the core.

The collapse stops as the density at the center approaches the nuclear density ($\sim 10^{15}$~g~cm$^{-3}$). At this time (referred to as ``core bounce"), the core with a radius of a few 10\,km consists of free neutrons with a small fraction of free protons (several percent), called a proto-neutron star. The gravitational energy due to the contraction of the core is immediately converted to thermal energy through shock heating, and finally almost entirely to energy in the form of neutrinos. These neutrinos, which are emitted from the neutrinosphere (located slightly below the proto-neutron star surface), are the driving sources of material outside the core, i.e., of the explosion. During the first 0.1--1~second, neutrino capture on free nucleons heats up the matter above the proto-neutron star surface and induces convective motion. This efficiently transports the neutrino energy to the material, leading to subsonic outflows called ``neutrino-heated ejecta". After the outer dense material is lifted, nearly freely streaming material from the neutrinosphere, principally due to neutrino capture on free nucleons, becomes transonic outflows lasting over 10 seconds, called ``neutrino-driven winds".

Historically, the innermost ejecta of core-collapse supernovae have been occasionally suggested as promising sites of r-process because of the expected  neutron richness in the vicinity of proto-neutron stars. Such early work includes a hypothetical scenario of the ``prompt explosion" from a low-mass progenitor ($\sim 8$--$10\, M_\odot$), in which very neutron-rich material near the proto-neutron star surface is immediately expelled by shock heating \citep{Hillebrandt1982}. However, recent hydrodynamical studies with elaborate treatment of neutrino transport have shown that such prompt explosions do not occur \citep{Hudepohl2010}. Thus, in the innermost ejecta of core-collapse supernovae, the matter is only moderately neutron-rich or even proton-rich, which appear to be unlikely sites of r-process nuclei but for iron-group and light trans-iron nuclei. In the following sections, nucleosynthesis in the neutrino-heated ejecta is outlined based on the latest results of multi-dimensional hydrodynamical simulations. Then, basic physics of neutrino-driven winds and resulting nucleosynthesis are discussed based on a well-established semi-analytic model. Finally, the main points of this chapter are summarized.

\section{\textit{Nucleosynthesis in neutrino-heated ejecta}}
\label{sec:ejecta}

\begin{figure}
    \centering
    \includegraphics[width=\textwidth]{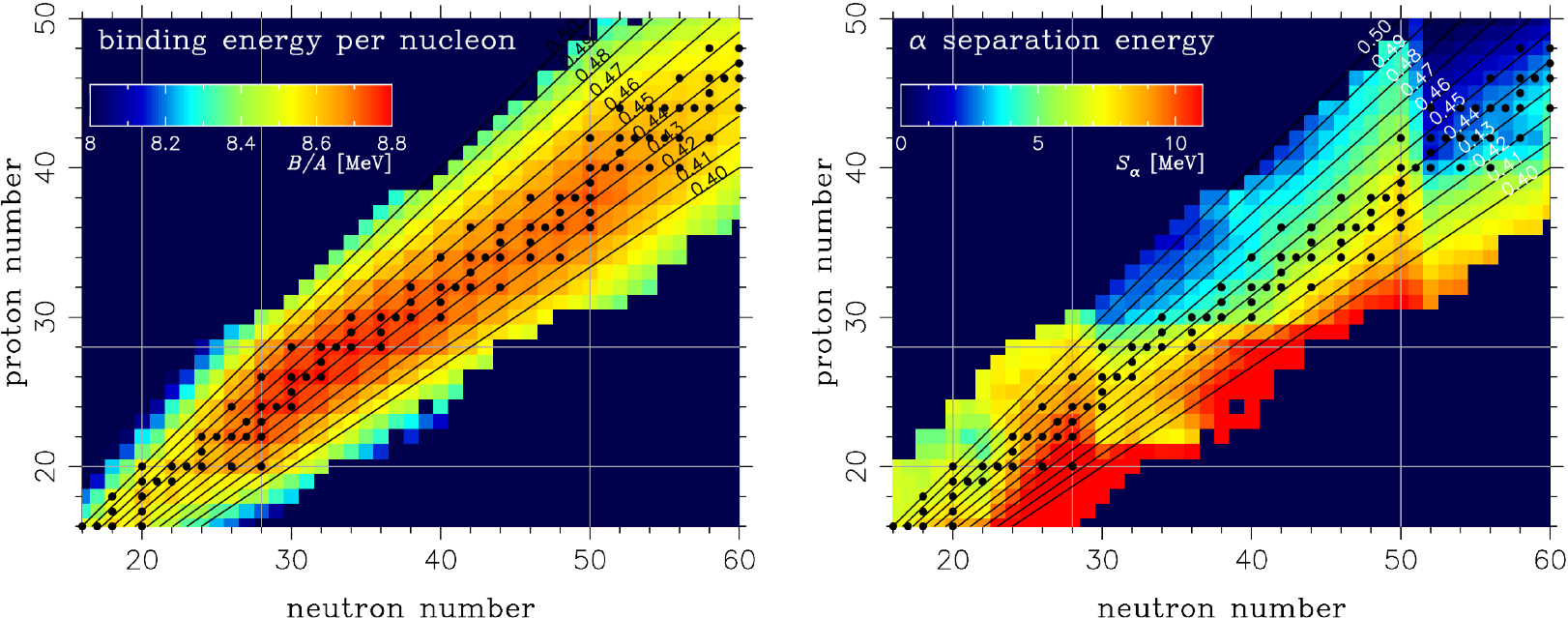}
    \caption{Binding energies per nucleon (left) and $\alpha$ separation energies (right) on the chart of nuclides in units of MeV for isotopes with experimentally evaluated masses \citep{Huang2017,Wang2021}. The circles denote stable isotopes. The black lines indicate the proton-to-nucleon ratios ($Y_\mathrm{e}$) for isotopes. The horizontal and vertical white lines are the magic numbers (20, 28, and 50) for protons and neutrons, respectively. (\citealt{Wanajo2018}; reproduced with permission of the AAS.)}
    \label{fig:basa}
\end{figure}

The explosion mechanism of core-collapse supernovae has not been fully understood (see, e.g., \citealt{Janka2012}). This makes the physical conditions relevant to nucleosynthesis in neutrino-heated ejecta somewhat uncertain, which are directly linked to the explosion mechanism itself. However, the mechanism for the low-mass end of progenitors ($\sim 8$--$10\, M_\odot$) is relatively well understood according to recent hydrodynamical studies \citep{Hudepohl2010,Melson2015,Wanajo2011,Wanajo2018}. For this reason, the model of a $9.6\, M_\odot$ star (labelled as z9.6 in \citealt{Wanajo2018}) is occasionally taken to be representative of core-collapse events with the aim of understanding basic pictures of nucleosynthesis in neutrino-heated ejecta. Dependences of nucleosynthetic outcomes on progenitor masses are then discussed in some detail.

\subsection{\textit{Nuclear statistical equilibrium (NSE)}}
\label{subsec:nse}

At the onset of core-collapse, the central core of a massive star, occasionally called an ``iron core", consists mainly of $^{56}$Ni. This is due to the fact that, at the temperature $\sim 5$~GK of the core (where GK~$= 10^9$~K), photodisintegration of nuclei releases an appreciable amount of free nucleons (neutrons and protons, labelled as $n$ and $p$). In such conditions, all nuclei are in nuclear statistical equilibrium (NSE), which are connected to each other through relevant reactions such as neutron capture, proton capture, $\alpha$ capture, and their inverse. That is, a given species with atomic number $Z$ and $A$, labelled as $(Z, A)$, is also in thermal equilibrium with free nucleons, 
\begin{equation}
    (A - Z)\, n + Z\, p \longleftrightarrow (Z, A) + \gamma,
    \label{eq:eq}
\end{equation}
where $\gamma$ indicates photons. Solving the equation of chemical equilibrium (i.e., the Saha equation) for Eq.~(\ref{eq:eq}) results in
\begin{equation}
    Y(Z, A) = \left(\frac{\rho}{m_\mathrm{u}}\right)^{A-1}
    \frac{G(Z, A, T)\, A^{3/2}}{2^A}
    (5.943\times 10^{33}\, T_\mathrm{GK}^{3/2})^{1-A}
    \exp \left(\frac{B(Z, A)}{kT}\right)
    Y_\mathrm{n}^{A-Z} Y_\mathrm{p}^Z,
    \label{eq:nse}
\end{equation}
where $Y_\mathrm{n}$, $Y_\mathrm{p}$, and $Y(Z, A)$ are, respectively, the numbers per nucleon (frequently called ``abundances") of neutrons, protons, and nuclei $(Z, A)$, $\rho$ is matter density, $T$ is temperature, $T_\mathrm{GK}$ is temperature in units of GK, $m_\mathrm{u}$ is atomic mass unit, $k$ is Boltzmann constant, and $G(Z, A, T)$ is partition function. $B(Z, A)$ is the binding energy of the species $(Z, A)$ defined as
\begin{equation}
    B(Z, A)/c^2 = (A-Z)\, m_\mathrm{n} + Z\, m_\mathrm{p} - M(Z,A),
    \label{eq:be}
\end{equation}
where $c$ is the speed of light and $m_\mathrm{n}$, $m_\mathrm{p}$, and $M(Z,A)$ are, respectively, the masses of a neutron, a proton, and a nucleus $(Z, A)$. Along with the equations of mass conservation
\begin{equation}
    \sum_{Z, A} X(Z, A) = A\, Y(Z, A) = 1
\end{equation}
and charge conservation
\begin{equation}
    \sum_{Z, A} Z\, Y(Z, A) = Y_\mathrm{e},
\end{equation}
$ Y(Z, A)$ can be computed for arbitrary $\rho$ and $T$. Here, $X(Z, A)$ is mass fraction and $Y_\mathrm{e}$ is the electron fraction. $Y_\mathrm{e}$ is equivalent to the number of protons (including those in nuclei) per nucleon for neutral material. Note that the core is made of nuclei with nearly equal numbers of neutrons and protons, i.e., $Y_\mathrm{e} \approx 0.5$. Because of the presence of $B(Z, A)$ in the exponent of Eq.~(\ref{eq:nse}), $^{56}$Ni becomes the most abundant species in the core, because it has the largest binding energy (or smallest mass; see Eq.~(\ref{eq:be})) among those nuclei. 

Once the core starts to collapse, the temperature rapidly increases owing to the core's nearly adiabatic contraction. As a result, photodisingegration of nuclei becomes progressively more efficient, leading to the dominance of free nucleons over nuclei (Eq.~(\ref{eq:eq})). The core matter is in NSE throughout the collapsing phase because the timescale to achieve NSE (within $\sim 1 \mu$s, \citealt{Meyer1998}) is shorter than the dynamical timescale (which is of the order of ms). Thus, this can be understood as a result of the reduction of the exponent in Eq.~(\ref{eq:nse}) due to increasing temperature. When the temperature exceeds $\sim 10$~GK, the core matter becomes entirely composed of free nucleons. During the collapse, the increasing degeneracy of electrons due to high density also induces electron capture on free protons as well as on nuclei. When the collapse stops, the core is composed of highly neutron rich material, $Y_\mathrm{e} < 0.1$, which is known as a proto-neutron star.

As the released gravitational energy is almost entirely converted to energy in the form of neutrinos, a copious amount of neutrinos begins to freely stream from the neutrinosphere beneath the proto-neutron star surface. This leads to outflows of material near the proto-neutron surface owing to neutrino heating. The heating is predominantly due to neutrino capture on free nucleons,
\begin{equation}
    \nu_\mathrm{e} + n \longrightarrow p + e^-
    \label{eq:nucap}
\end{equation}
and
\begin{equation}
    \bar{\nu}_\mathrm{e} + p \longrightarrow n + e^+,
    \label{eq:anucap}
\end{equation}
where $\nu_\mathrm{e}$, $\bar{\nu}_\mathrm{e}$, $e^-$, and $e^+$ are, respectively, electron neutrino, electron anti-neutrino, electron, and positron. Therefore, $Y_\mathrm{e}$, which is equivalent to the number of protons per nucleon in free-nucleon dominated material, inevitably changes in the ejecta. Since these reactions occur almost equally, the number of protons and neutrons becomes similar. Thus, $Y_\mathrm{e}$ increases from an initially small value ($< 0.1$) to $\sim 0.5$. Recent studies indicate the resulting range of $Y_\mathrm{e} \sim 0.4$--0.6, which depends on the progenitor masses \citep{Wanajo2018}.

The temperature decreases from the initial values of a few 10~GK as the outflowing matter expands. The matter is still in NSE with the abundance distribution determined by Eq.~(\ref{eq:nse}) until the temperature decreases to $\sim 6$~GK. Thus, free nucleons reassemble to iron-group nuclei along with decreasing temperature.

\begin{figure}
    \centering
    \includegraphics[width=1.08\textwidth]{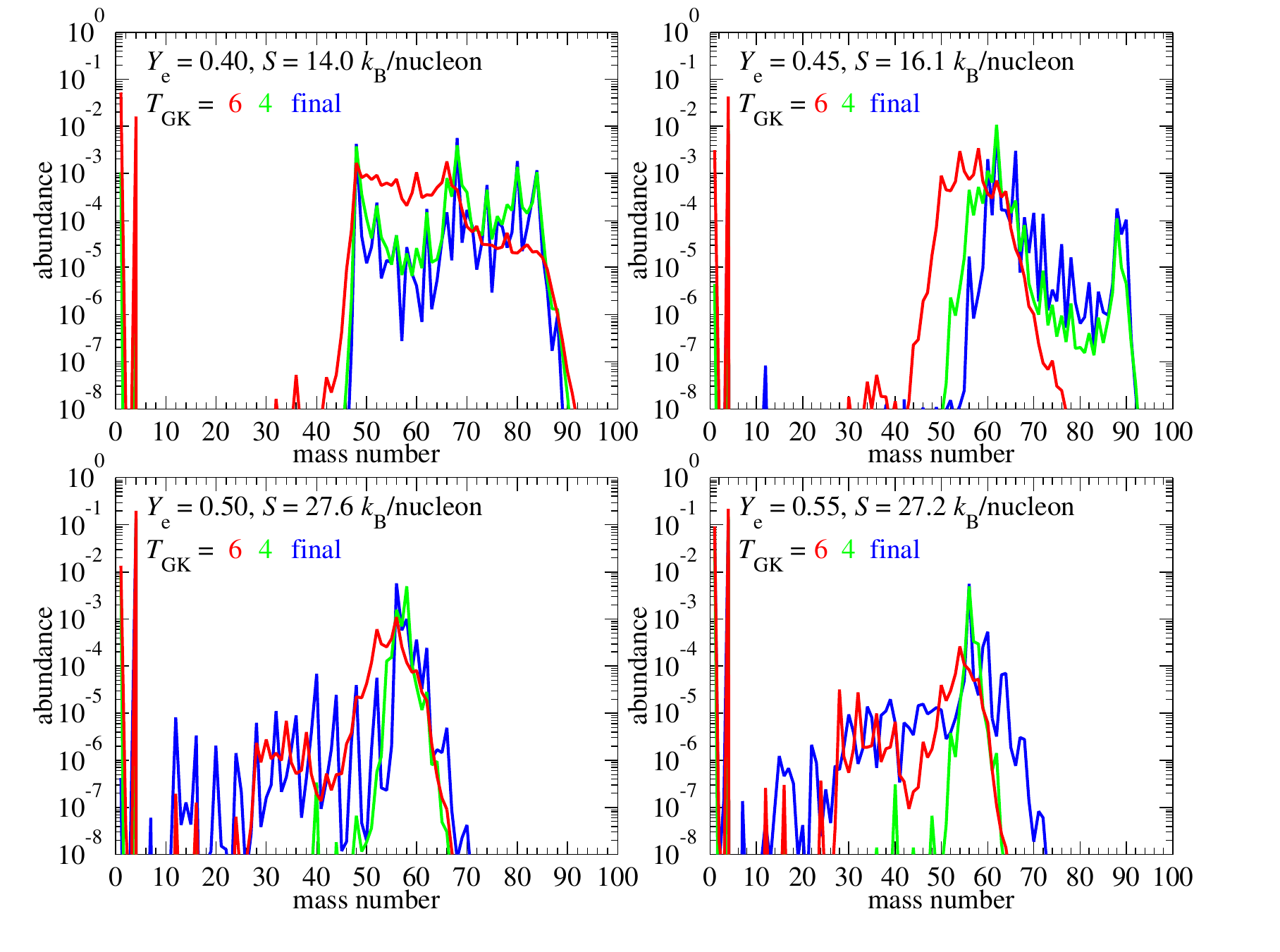}
    \caption{Abundances with decreasing temperature at $T_\mathrm{GK} = 6$ and 4 as well as those at the end of calculations (``final" in panels) for a $9.6\, M_\odot$ model (z9.6 in \citealt{Wanajo2018}). Selected trajectories are those with initial $Y_\mathrm{e} = 0.40$ (left top), 0.45 (right top), 0.50 (left bottom), and 0.55 (right bottom). The entropy per nucleon, $S$, in units of Boltzmann constant $k_\mathrm{B}$ is also shown in the legend of each panel. (\citealt{Wanajo2018}; reproduced with permission of the AAS.)}
    \label{fig:freeze}
\end{figure}

During the NSE phase ($T_\mathrm{GK} > 6$), the two groups of light particles ($n$, $p$, and $\alpha$) and heavy nuclei ($Z > 2$) are causally connected by relevant reactions, that is, all nuclei belong to a single NSE cluster. Two sub-clusters consisting of light particles and heavy nuclei, respectively, are in thermal equilibrium through three-body reactions and their inverse,
\begin{equation}
    \alpha + \alpha + \alpha \longleftrightarrow ^{12}\mathrm{C} + \gamma
    \label{eq:3a}
\end{equation}
and
\begin{equation}
    \alpha + \alpha + n \longleftrightarrow ^9\mathrm{Be} + \gamma,
    \label{eq:aan}
\end{equation}
the latter followed by
\begin{equation}
    ^9\mathrm{Be} + \alpha \longleftrightarrow ^{12}\mathrm{C} + n.
    \label{eq:bean}
\end{equation}
The forward reactions in Eqs.~(\ref{eq:3a}) and (\ref{eq:aan}) act as  channels from light to heavy nuclei for $Y_\mathrm{e} > 0.49$ (only slightly neutron-rich or proton-rich) and $Y_\mathrm{e} < 0.49$ (neutron-rich), respectively. As the matter expands, these three-body reactions progressively slow down (owing to their rates being proportional to $\rho^2$) and finally become inefficient when temperature decreases to $T_\mathrm{GK} \sim 6$ \citep{Meyer1998,Wanajo2013}. This is the end of the NSE phase.

Fig.~\ref{fig:freeze} shows the abundance patterns for the selected thermodynamic trajectories (with different initial $Y_\mathrm{e}$) of a $9.6\, M_\odot$ model (z9.6 in \citealt{Wanajo2018}) as representative. The dependencies of nucleosynthetic outcomes on progenitor masses will be discussed in the subsequent subsection. At $T_\mathrm{GK} = 6$ (approximately the end of the NSE phase), the most abundant isotope (except for $n$, $p$, and $\alpha$) is the one having the greatest binding energy near a given initial $Y_\mathrm{e}$ on the chart of nuclides (Fig.~\ref{fig:basa}, left): e.g., $^{48}$Ca and $^{56}$Ni for $Y_\mathrm{e} = 0.40$ and 0.50, respectively. Note that the isotopes on the neutron magic number $N = 20$, 28, and 50 as well as on the proton magic number $Z = 20$ and 28 (Fig.~\ref{fig:basa}, left), called magic number nuclei, have large binding energies. Both $^{48}$Ca and $^{56}$Ni are double magic number nuclei with $(Z, N) = (20, 28)$ and $(Z, N) = (28, 28)$, respectively.

\subsection{\textit{Nuclear statistical quasi-equilibrium (QSE)}}
\label{subsec:qse}

After the end of the NSE phase ($T_\mathrm{GK} \sim 6$), each group consisting of light or heavy nuclei is still in thermal equilibrium, a state known as nuclear statistical \textit{quasi}-equilibrium (QSE), because of two-body reactions (the rates being proportional to $\rho$) faster than three-body reactions. That is, the single NSE cluster splits into two separate QSE clusters. Light particle capture on heavy nuclei and its inverse are still active at this stage, while the gateway between two QSE clusters, Eqs.~(\ref{eq:3a}) and (\ref{eq:aan}) is closed, i.e.,
\begin{equation}
    Y_\mathrm{light} = \sum_{Z \le 2} Y(Z, A) \ne \mathrm{const.}
    \quad \mathrm{and} \quad
    Y_\mathrm{heavy} = \sum_{Z > 2} Y(Z, A) = \mathrm{const.}
    \label{eq:qse}
\end{equation}
Under these conditions, the subsequent evolution of heavy abundance distributions are governed by the amounts of light particles, in particular of $\alpha$ particles \citep{Meyer1998}.

\begin{figure}
    \centering
    \includegraphics[width=1.08\textwidth]{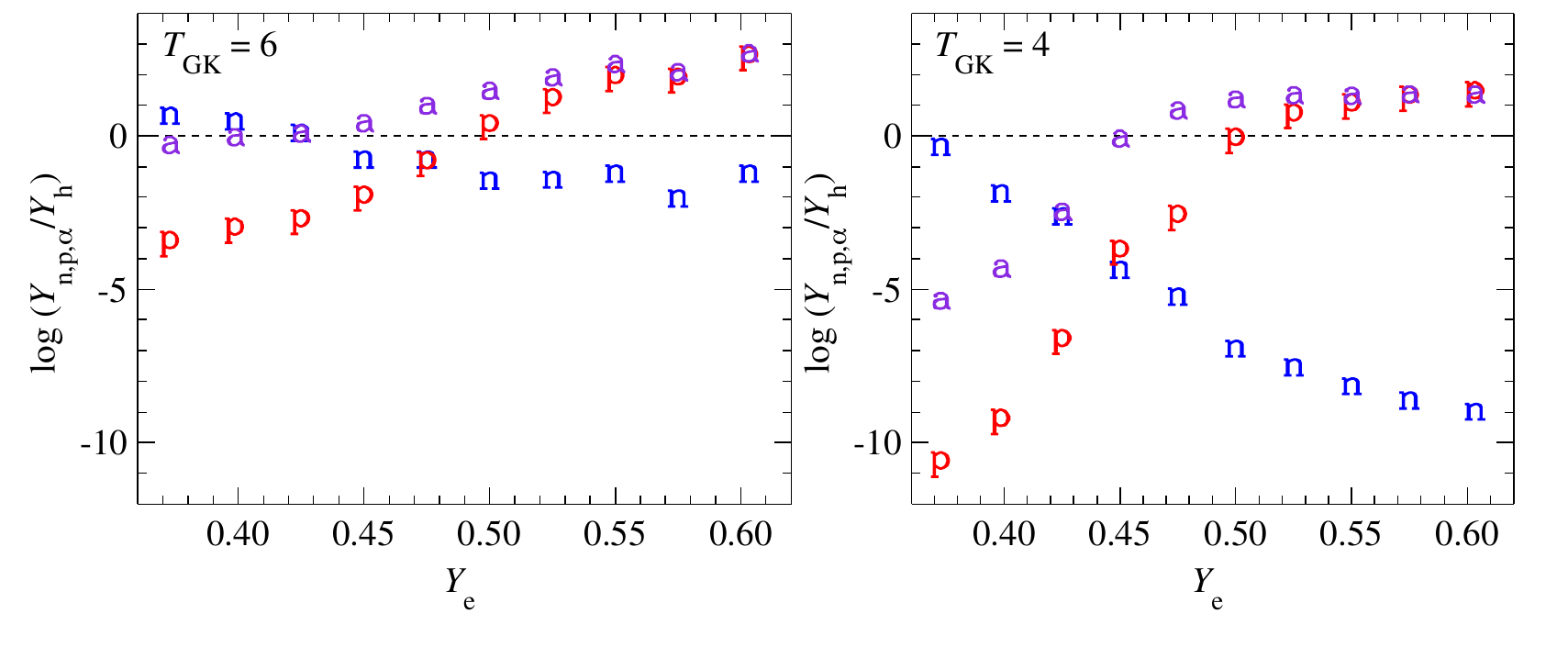}
    \caption{Abundances of neutrons (``n"), protons (``p"), and $\alpha$ particles (``a") with respect to those of heavy nuclei ($Z > 2$) at $T_\mathrm{GK} = 6$ (left) and 4 (right) for selected trajectories of a $9.6\, M_\odot$ model (z9.6 in \citealt{Wanajo2018}). The dashed line in each panel indicates a ratio of unity. (\citealt{Wanajo2018}; reproduced with permission of the AAS.)}
    \label{fig:toseed}
\end{figure}

Fig.~\ref{fig:toseed} (left) shows the abundances of $n$, $p$, and $\alpha$ with respect to those of heavy nuclei ($Z > 2$) at the end of NSE ($T_\mathrm{GK} = 6$) for selected trajectories of model z9.6. At this time, $\alpha$ particles are less and more abundant with respect to heavy nuclei for $Y_\mathrm{e} < 0.43$ and $Y_\mathrm{e} > 0.43$, respectively, referred to as $\alpha$-deficient QSE and $\alpha$-rich QSE hereafter \citep{Meyer1998,Wanajo2013}. During the QSE phase ($T_\mathrm{GK} \sim 6$--4), the abundance distribution reshapes so that the isotopes with greater $\alpha$-separation energies (Fig.~\ref{fig:basa}, right) become more abundant. Here, the $\alpha$-separation energy, $S_\alpha(Z, A)$, of the isotope $(Z, A)$ is defined as
\begin{equation}
    S_\alpha(Z, A)/c^2 = M(Z-2,A-4) + m_\alpha - M(Z,A),
    \label{eq:be}
\end{equation}
where $m_\alpha$ is the mass of an $\alpha$ particle. 

For $\alpha$-deficient QSE, e.g., $Y_\mathrm{e} = 0.40$ in the left top of Fig.~\ref{fig:freeze}, the number of $\alpha$ particles is too small to substantially bring nuclear abundances beyond $A = 68$, resulting in the most abundant isotopes of $^{48}$Ca and $^{68}$Ni along with prominent production of some light trans-iron species such as $^{82, 84}$Se at the end of the QSE phase ($T_\mathrm{GK} = 4$). For $\alpha$-rich QSE in slightly neutron-rich conditions, e.g., $Y_\mathrm{e} = 0.45$ in the right top of Fig.~\ref{fig:freeze}, the number of $\alpha$ particles is sufficiently large such that appreciable nuclear abundances of $A \sim 50$--60 are brought to $A \sim 90$ on the magic number $N = 50$ such as $^{88}$Sr, $^{89}$Y, and $^{90}$Zr, while those of $A \sim 60$--90 are deficient. In this way, light trans-iron nuclei up to $A \sim 90$ are produced in neutron-rich conditions during the QSE phase. It is important to note that a proton-rich isotope $^{92}$Mo (whose origin is currently unknown) on $N = 50$ is also produced in $\alpha$-rich QSE with $Y_\mathrm{e} \sim 0.47$ \citep{Wanajo2018}.

For $\alpha$-rich QSE in symmetric and proton-rich conditions, e.g., $Y_\mathrm{e} = 0.50$ and $Y_\mathrm{e} = 0.55$ in the left bottom and right bottom of Fig.~\ref{fig:freeze}, respectively, the most abundant abundances at the end of the QSE phase are $^{56}$Ni with $N = Z = 28$. There are few heavier isotopes to be made with sufficiently large $\alpha$-separation energies at $Y_\mathrm{e} \ge 0.50$ as can be seen in the right panel of Fig.~\ref{fig:basa}.

\subsection{\textit{(No) $r$-process, $\alpha$-process, and $\nu$p-process}}
\label{subsec:process}

The QSE phase ends as the temperature decreases to $T_\mathrm{GK} \sim 4$ at which the photodisintegration of nuclei releasing $\alpha$ particles becomes inefficient. At this time, free neutrons are substantially under-abundant with respect to heavy nuclei for the entire range of $Y_\mathrm{e}$ as can be seen in the right panel of Fig.~\ref{fig:toseed}. That is, the neutrino-heated ejecta do not provide the physical conditions required for the $r$-process. Similarly, $\alpha$ particles are under-abundant with respect to heavy nuclei for $Y_\mathrm{e} < 0.45$, leading to only slight modifications in abundance distributions by subsequent $\alpha$ capture (see small differences between green and blue curves in the top panels of Fig.~\ref{fig:freeze}).

\begin{figure}
    \centering
    \includegraphics[width=0.8\textwidth]{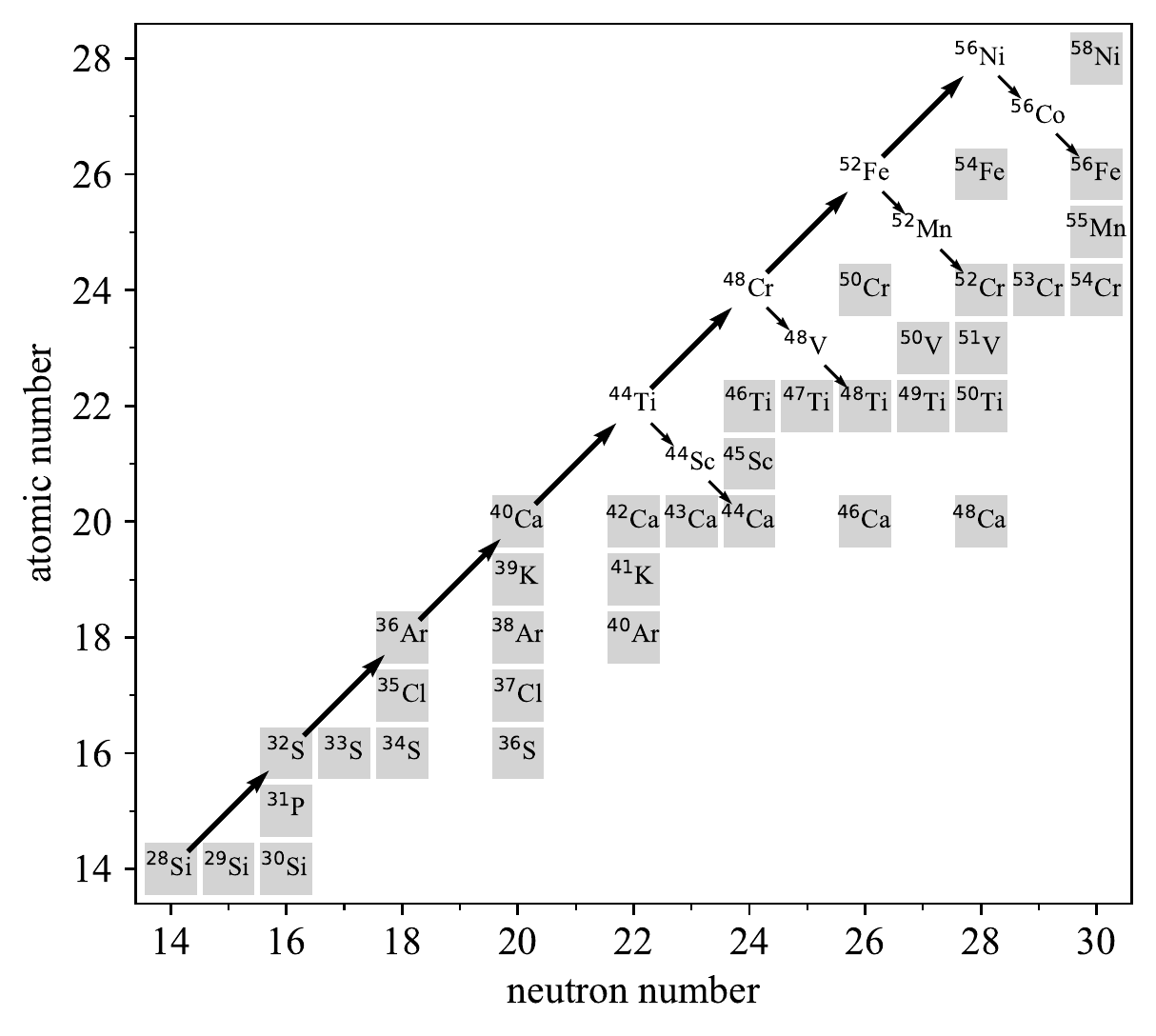}
    \caption{Main paths of an $\alpha$-process shown by thick arrows. Thin arrows denote the subsequent decays (predominantly by electron capture) toward $\beta$-stability. Gray squares indicate stable isotopes.}
    \label{fig:alpha}
\end{figure}

For $Y_\mathrm{e} > 0.45$, $\alpha$ particles are more abundant than heavy nuclei at the end of QSE phase (Fig.~\ref{fig:toseed}, right), called ``$\alpha$-rich freeze-out". In such conditions, the triple-$\alpha$ (forward reaction in Eq.~(\ref{eq:3a})) produces $^{12}$C and heavier $\alpha$ elements, which is referred to as ``$\alpha$-process" \citep{Woosley1992}. Fig.~\ref{fig:alpha} illustrates the path of an $\alpha$-process approaching $^{56}$Ni. As can be seen in the left-bottom panel of Fig.~\ref{fig:freeze}, an $\alpha$-process appreciably enhances the amounts of $\alpha$-elements (with $A$ multiple of 4), including a long-lived isotope $^{44}$Ti (half-life of 60~yr; progenitor of stable $^{44}$Ca). Nuclei heavier than $^{56}$Ni, such as $^{64}$Ge (progenitor of stable $^{64}$Zn), are not substantially enhanced owing to their increasing height of Coulomb barrier.

\begin{figure}
    \centering
    \includegraphics[width=0.6\textwidth]{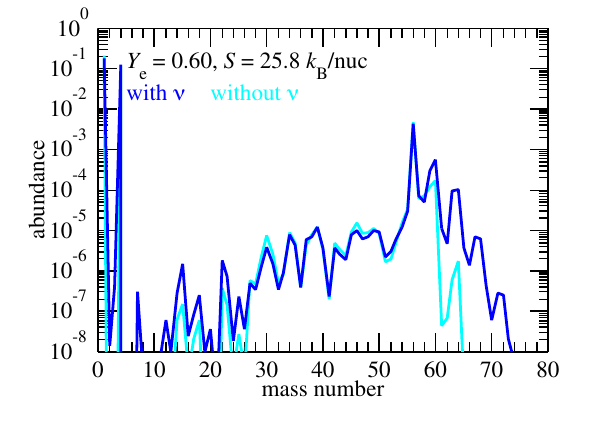}
    \caption{Final abundances with (blue) and without (cyan) neutrino capture on free nucleons for a $9.6\, M_\odot$ model (z9.6 in \citealt{Wanajo2018}) with the trajectory of $Y_\mathrm{e} = 0.60$. (\citealt{Wanajo2018}; reproduced with permission of the AAS.)}
    \label{fig:nuonoff}
\end{figure}

For $Y_\mathrm{e} > 0.5$, free protons are also more abundant than heavy nuclei at the end of QSE phase (Fig.~\ref{fig:toseed}, right). Because of the lower Coulomb barrier compared to that for $\alpha$ particles, proton capture,
\begin{equation}
    (Z, A) + p \longrightarrow (Z+1, A+1) + \gamma,
    \label{eq:pg}
\end{equation}
starting from $^{56}$Ni plays a role for producing some species beyond iron. Here, $\beta^+$-decay (e.g., of $^{64}$Ge with half-life of 1 min) after each proton capture can be replaced by the much faster neutron capture,
\begin{equation}
    (Z+1, A+1) + n \longrightarrow (Z, A+1) + p,
    \label{eq:np}
\end{equation}
by consuming free neutrons released from neutrino capture on free protons (Eq.~(\ref{eq:anucap})). A chain of these successive proton and neutron captures in the ejecta subject to intense neutrino irradiation is called a ``$\nu$p-process" \citep{Froehlich2006,Pruet2006,Wanajo2006}. Fig.~\ref{fig:nuonoff} indicates that a $\nu$p-process (with neutrino capture on free nucleons) appreciably enhances the abundances of $A \sim 60$--70. The $\nu$p-process appears to play only a subdominant role in the neutrino-heated ejecta (Fig.~\ref{fig:freeze}, left bottom) except for, e.g., producing $^{64}$Zn (made as $^{64}$Ge; whose origin is currently unknown). However, this process can be the dominant sources of some proton-rich isotopes up to $A \sim 110$ \citep{Wanajo2011b} as detailed in the next section.

\subsection{\textit{Dependence on progenitor masses}}
\label{subsec:mass}

The basic picture of the nucleosynthetic history in neutrino-heated ejecta, starting from NSE to QSE and resulting in non-equilibrium light-particle capture processes along with decreasing temperature, is similar between explosions with different progenitor masses. However, the distributions of synthesized nuclear abundances are highly dependent on the progenitor masses. In this subsection, the various nucleosynthetic outcomes from the two-dimensional models with different progenitor masses are outlined, focusing on those near the low-mass end ($8.8\, M_\odot$; model e8.8 and $9.6\, M_\odot$; z9.6 in \citealt{Wanajo2018}) and heavier ($15\, M_\odot$; s15 and $27\, M_\odot$; s27 in \citealt{Wanajo2018}). The former and the latter can be regarded as representative of, respectively, low-mass and typical-mass ($15\, M_\odot$) or massive ($27\, M_\odot$) progenitors. Here, the progenitors are assumed to be non-rotating, non-magnetized, and non-binary stars (although these effects can be crucial in some cases).

\begin{figure}
    \centering
    \includegraphics[width=0.8\textwidth]{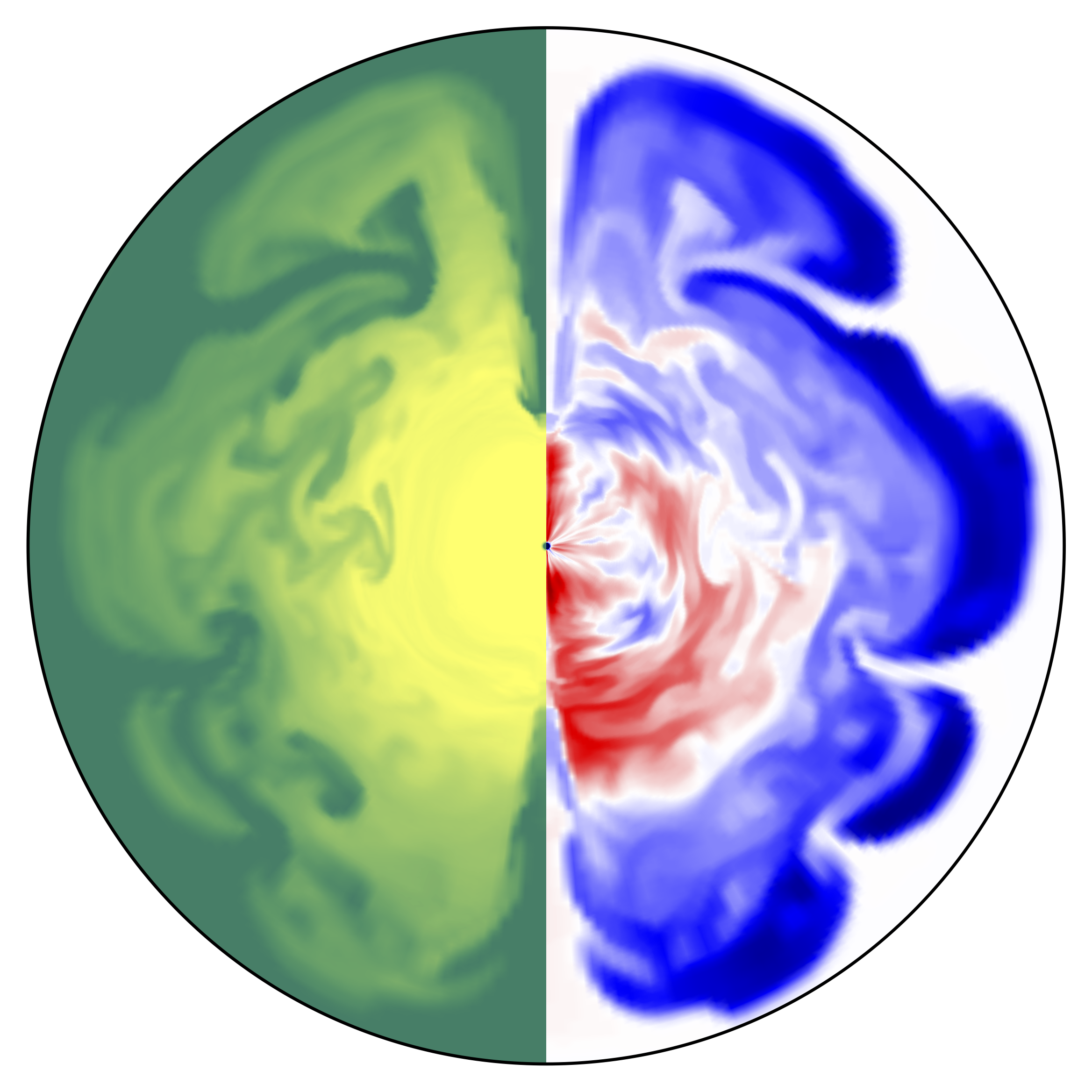}
    \caption{Snapshot of an $8.8\, M_\odot$ model (electron-capture supernova; model e8.8 in \citealt{Wanajo2018}) at 365~ms after core bounce in the central region (radius of 7200~km) of the exploding star. The left-half of the circle shows the entropy distribution with greenish and yellowish colors for lower and higher values, respectively. The right-half of the circle shows the distribution of electron fraction, $Y_\mathrm{e}$,  with bluish and reddish colors for neutron-rich ($Y_\mathrm{e} < 0.5$) and proton-rich ($Y_\mathrm{e} > 0.5$) regions, respectively. White indicates the regions with $Y_\mathrm{e} \approx 0.5$. (\citealt{Wanajo2018}; reproduced with permission of the AAS.) }
    \label{fig:ecsn}
\end{figure}

Fig.~\ref{fig:ecsn} depicts a snapshot taken from a core-collapse simulation of an $8.8\, M_\odot$ star at 365~ms after core bounce, in which entropy ($S$; green to yellow for low to high values) and $Y_\mathrm{e}$ (blue and red for neutron-rich and proton-rich regions, respectively) are shown in the left and right panels, respectively. The core of this progenitor star in the asymptotic-giant branch consists predominantly of O, Ne, and Mg (not $^{56}$Ni as in the case of more massive progenitors). The collapse is due to electron capture on some intermediate-mass nuclei at the center. The contraction of material owing to the loss of electron-degeneracy pressure at the center leads to a temperature increase sufficient for oxygen burning, which immediately propagates toward the core surface. At this time, the core chiefly consists of $^{56}$Ni; the subsequent fate is similar to that of the core-collapse supernovae from an iron core, which is known as an electron-capture supernova \citep{Nomoto1987}.

\begin{figure}
    \centering
    \includegraphics[width=1.08\textwidth]{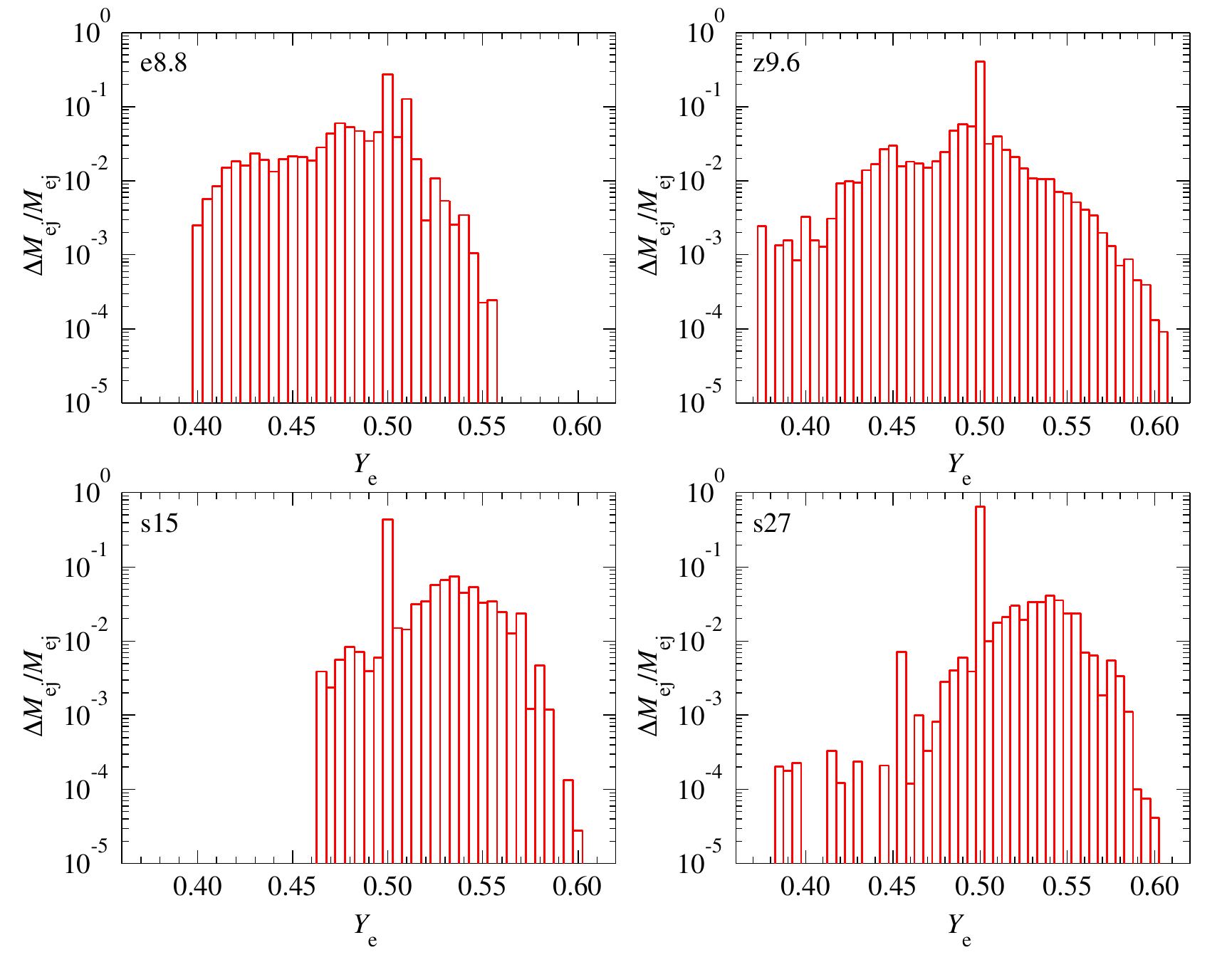}
    \caption{$Y_\mathrm{e}$ histograms for ejecta from progenitors of $8.8\, M_\odot$ (left top; model e8.8 in \citealt{Wanajo2018}), $9.6\, M_\odot$ (right top; z9.6), $15\, M_\odot$ (left bottom; s15), and $27\, M_\odot$ (right bottom; s27). Each panel shows the masses ($\Delta M_\mathrm{ej}$) with respect to the total ejecta mass ($M_\mathrm{ej}$) in $Y_\mathrm{e}$ bins with a width of $\Delta Y_\mathrm{e} = 0.005$. The spike at $Y_\mathrm{e} \approx 0.50$ indicates the material relatively distant from the core, which is not neutrino-processed and keeps the initial values of $Y_\mathrm{e}$. (\citealt{Wanajo2018}; reproduced with permission of the AAS.)}
    \label{fig:yehist}
\end{figure}

\begin{figure}
    \centering
    \includegraphics[width=1.08\textwidth]{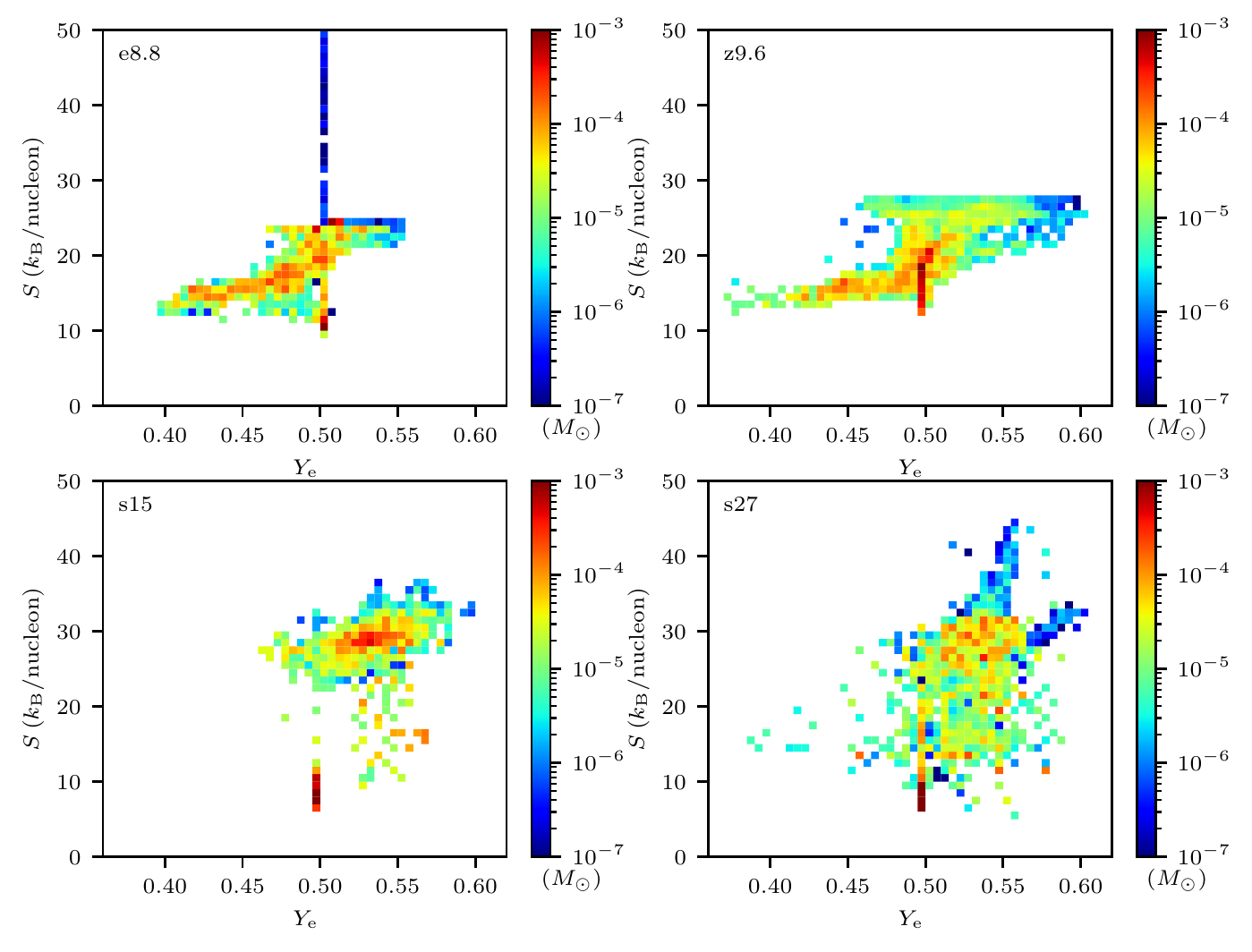}
    \caption{Distributions of ejecta masses on $Y_\mathrm{e}$--$S$ plane for progenitors of $8.8\, M_\odot$ (left top; model e8.8 in \citealt{Wanajo2018}), $9.6\, M_\odot$ (right top; z9.6), $15\, M_\odot$ (left bottom; s15), and $27\, M_\odot$ (right bottom; s27). (\citealt{Wanajo2018}; reproduced with permission of the AAS.)}
    \label{fig:yes}
\end{figure}

Earlier ejecta (at larger distance from the center) exhibit relatively lower $S$ and $Y_\mathrm{e}$, which indicate modest neutrino heating by neutrino capture on free nucleons in Eqs.~(\ref{eq:nucap}) and (\ref{eq:anucap}). This is due to the rapid outgoing material that is marginally decelerated by the diluted outer-core envelope (consisting of He and H) of an asymptotic-giant-branch star. By contrast, later ejecta which are subject to more neutrino heating have higher $S$ and $Y_\mathrm{e}$. For these reasons, the values of $Y_\mathrm{e}$ span a wide range of $\sim 0.4$--0.55 as can be seen in the left-top panel of Fig.~\ref{fig:yehist}. A clear correlation between the values of $Y_\mathrm{e}$ and $S$ is also found in the left-top panel of Fig.~\ref{fig:yes}, a consequence of the fact that neutrino heating predominantly due to the reactions in Eqs.~(\ref{eq:nucap}) and (\ref{eq:anucap}) raises both these quantities.

The model of $9.6\, M_\odot$ (nearly at the low-mass end for iron-core progenitors) exhibits similar behavior to that of $8.8\, M_\odot$ because of thin O and C shells between the core and the low-density He and H envelope. Namely, the rapidly expanding ejecta lead to the presence of a substantial amount of neutron-rich ($Y_\mathrm{e} \sim 0.4$) and low $S$ ($\sim 13$--$15\, k_\mathrm{B}$/nucleon) material. However, for more massive progenitors, i.e., $15\, M_\odot$ and $27\, M_\odot$ stars, the dense outer material above the core inhibits rapid ejecta expansion. This induces long-term convective activity in the ejecta between the core and the outer dense material, resulting in appreciable neutrino heating and thus higher $S$ and $Y_\mathrm{e}$. Fig.~\ref{fig:yehist} indicates the resultant deficiency of low $Y_\mathrm{e} (< 0.45)$ material as well as the peak of the distribution located in the proton-rich side ($Y_\mathrm{e} = 0.53$--0.54, except for the spike at $Y_\mathrm{e} \approx 0.5$; see caption). Fig.~\ref{fig:yes} also shows the dominance of higher $S$ ($\sim 30\, k_\mathrm{B}$/nucleon) material in the ejecta for these massive models. 

\subsection{\textit{Contribution to galactic as well as solar-system abundances}}
\label{subsec:solar}

\begin{figure}
    \centering
    \includegraphics[width=1.08\textwidth]{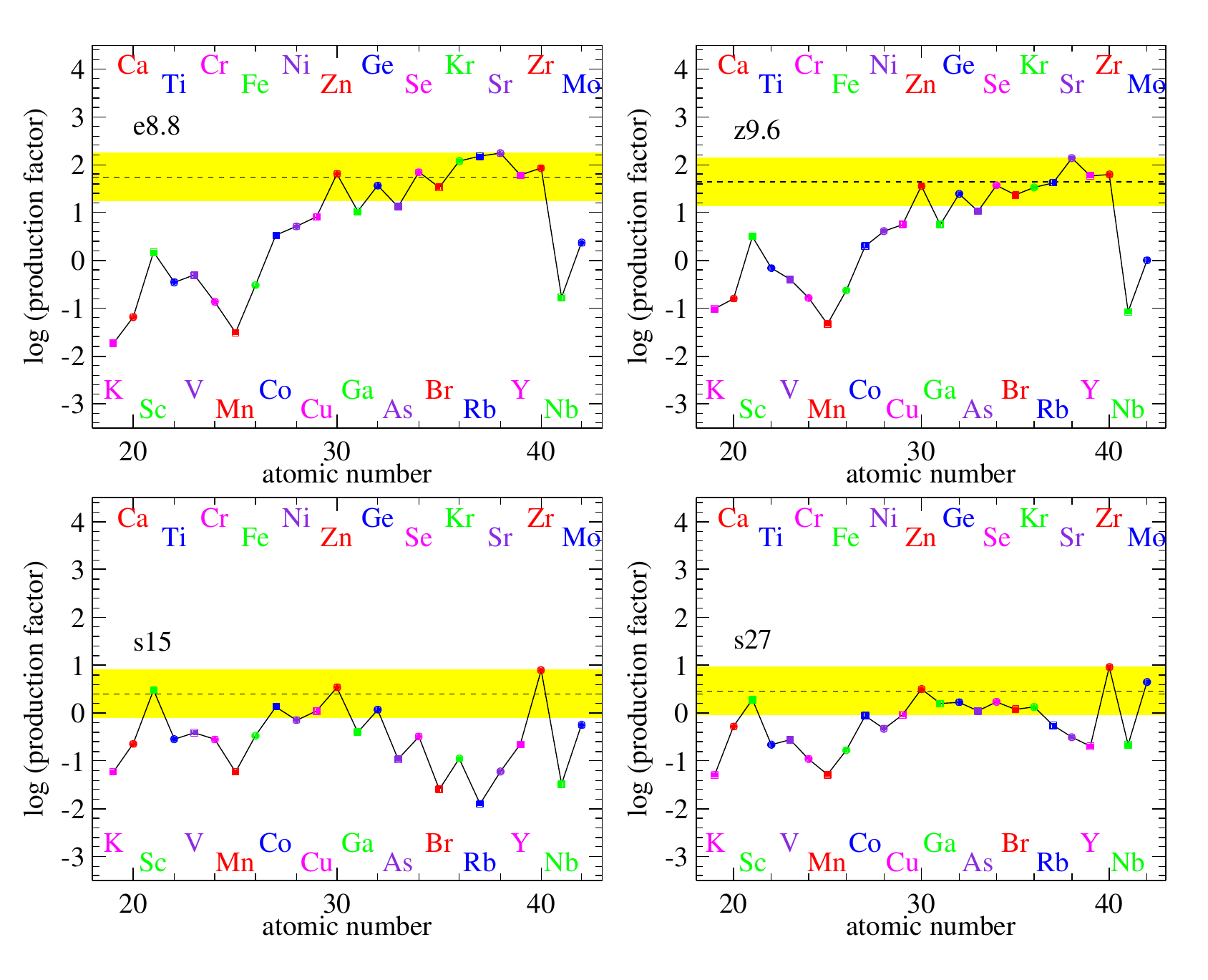}
    \caption{Elemental abundances in the total ejecta with respect to their solar values \citep{Lodders2009} for progenitors of $8.8\, M_\odot$ (left top; model e8.8 in \citealt{Wanajo2018}), $9.6\, M_\odot$ (right top; z9.6), $15\, M_\odot$ (left bottom; s15), and $27\, M_\odot$ (right bottom; s27). In each panel the normalization band defined as the range between the maximum value and one tenth of that is indicated in yellow with the median value (dashed line). (\citealt{Wanajo2018}; reproduced with permission of the AAS.)}
    \label{fig:pfel}
\end{figure}

Nucleosynthetic abundances with respect to their solar values \citep{Lodders2009}, called ``production factors", are shown as functions of atomic number in Fig.~\ref{fig:pfel} for all models. Elements that reside on the yellow ``normalization band", defined as the range between the maximum value and one tenth of that, can be regarded as candidates that contribute to the solar abundances. Note that the core-collapse simulation of each model here considers only the inner-most region of a progenitor star, that is, the central core and a fraction of the outer layer (He and Si shells for the stars of $8.8\, M_\odot$ and the others, respectively). The contributions (for, e.g., $\alpha$ and iron-group elements) from the outer layers are expected to be subdominant for the low-mass models (of $8.8\, M_\odot$ and $9.6\, M_\odot$) but not for the typical-mass or massive models (of $15\, M_\odot$ and $27\, M_\odot$). Moreover, for the latter, the convective motion was still active at the end of the simulation and thus the mass ejection from the inner-most region had not been finalized. Thus, the amounts of synthesized elements for the massive models should be taken as lower limits.

According to the studies of  galactic chemical evolution (e.g., \citealt{Prantzos2018}), the requisite production factors are $\sim 10$ so as ``typical" core-collapse supernovae to be the dominant contributors of elements. That is, the overall large production factors ($\sim 100$) of $Z = 30$--40 (from Zn to Zr) for the models of $8.8\, M_\odot$ and $9.6\, M_\odot$ indicate that these elements can originate from such low-mass core-collapse supernovae if their progenitors account for $\sim 10\%$ of all core-collapse supernovae. This corresponds to the progenitor mass window of $\sim 1\, M_\odot$, which is currently uncertain (some studies indicate the mass window of only $\sim 0.2\, M_\odot$ for electron capture supernovae, \citealt{Poelarends2008,Doherty2015}). For more massive cases ($15\, M_\odot$ and $27\, M_\odot$), the overall production factors are below $\sim 10$ (bottom panels of Fig.~\ref{fig:pfel}). However, recalling that simulations of these models ended during mass ejection, it seems possible that some elements such as Sc, Zn, Zr, and Mo can, in part, originate from massive core-collapse supernovae.

\begin{figure}
    \centering
    \includegraphics[width=1.08\textwidth]{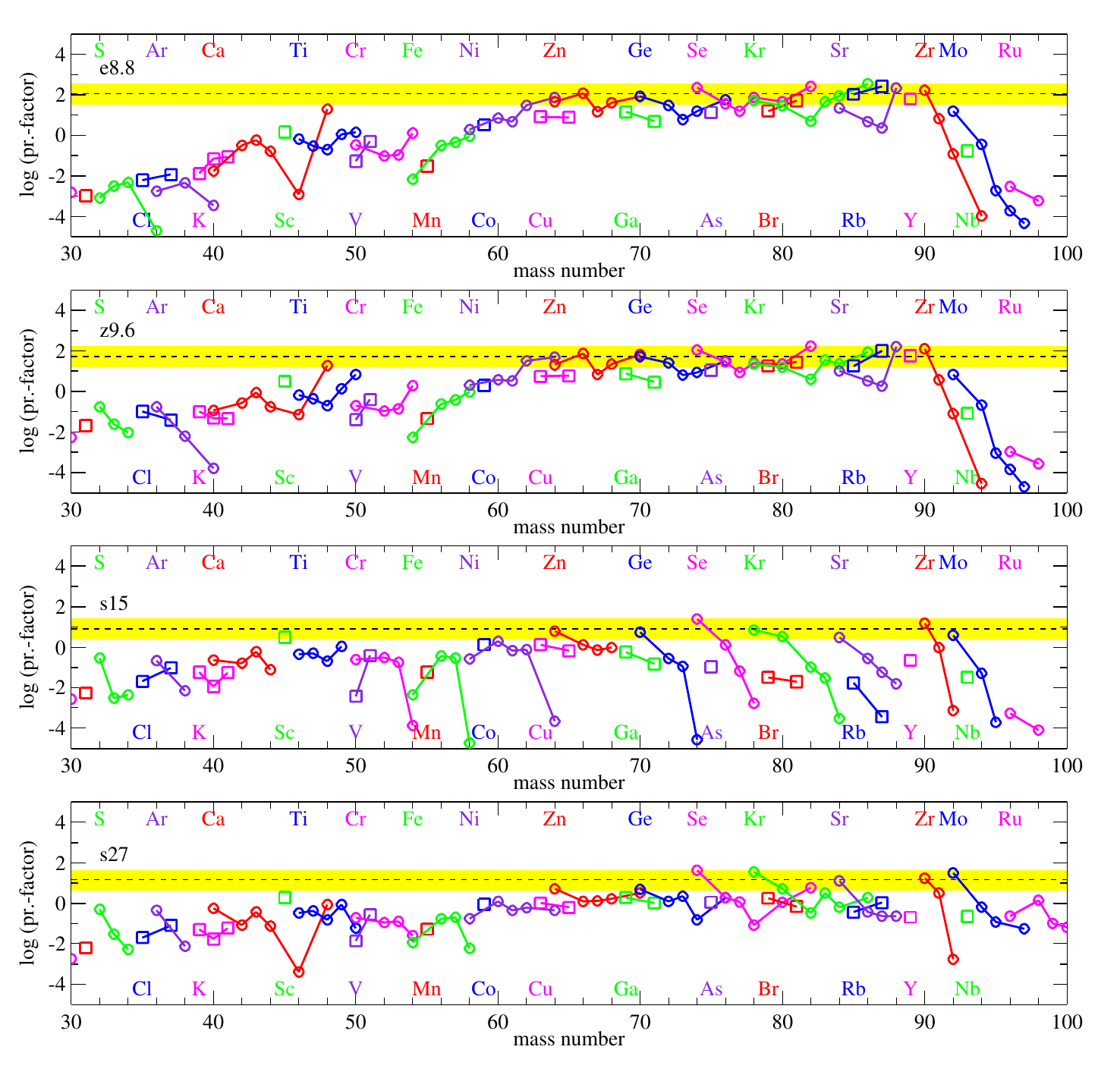}
    \caption{Isotopic abundances in the total ejecta with respect to their solar values \citep{Lodders2009} for progenitors of $8.8\, M_\odot$ (left top; model e8.8 in \citealt{Wanajo2018}), $9.6\, M_\odot$ (right top; z9.6), $15\, M_\odot$ (left bottom; s15), and $27\, M_\odot$ (right bottom; s27). In each panel the normalization band defined as the range between the maximum value and one tenth of that is indicated in yellow with the median value (dashed line). (\citealt{Wanajo2018}; reproduced with permission of the AAS.)}
    \label{fig:pfiso}
\end{figure}

Plots of the production factors of isotopes are convenient for more quantitative analysis, which are shown in Fig.~\ref{fig:pfiso} as functions of mass number for all models. For low-mass models ($8.8\, M_\odot$ and $9.6\, M_\odot$), the production factors of the majority of isotopes with $A = 64$--90 exhibit a flat trend (1st and 2nd panels of Fig.~\ref{fig:pfiso}), residing on the normalization band. This indicates that low-mass core-collapse supernovae are promising sources of elements with $Z = 30$--40 (from Zn to Zr) including their isotopes. It is important to note that these light trans-iron isotopes, frequently categorized as light s-process or r-process nuclei, are produced in neutron-rich ejecta ($Y_\mathrm{e} < 0.5$) of these models during the QSE phase and by a subsequent $\alpha$-process (top panels of Fig.~\ref{fig:freeze}), not by neutron-capture processes. 

For this reason, such a flat trend of production factors cannot be seen in more massive models ($15\, M_\odot$ and $27\, M_\odot$). Instead, appreciable production factors of $\sim 10$--40 are found for the proton-rich isotopes of elements with $Z = 30$--42 (from Zn to Mo), which are mainly produced in the ejecta with $Y_\mathrm{e} > 0.45$--0.5 (i.e., sufficiently proton-rich compared to the mean of $\beta$-stability; the left panel of Fig.~\ref{fig:basa}) during the QSE phase and in part by a subsequent $\nu$p-process. Some of these isotopes ($^{74}$Se, $^{78}$Kr, $^{84}$Sr, and $^{92}$Mo) are so-called ``p-nuclei", among which in particular the origin of $^{92}$Mo is currently unknown. These plots (3rd and 4th of Fig.~\ref{fig:pfiso}) imply that typical-mass or massive core-collapse supernovae can be predominant sources of these light p-nuclei including $^{92}$Mo. Here, $^{92}$Mo is produced during the QSE phase and by subsequent proton capture on the magic number $N = 50$ in slightly neutron-rich environment ($Y_\mathrm{e} \sim 0.47$, \citealt{Hoffman1996,Wanajo2018}).

It is noteworthy that $^{64}$Zn, the main isotope of Zn, is appreciably produced in all models presented here, for which the astrophysical origin is currently unknown. For the low-mass models, $^{64}$Zn is exclusively synthesized in neutron-rich ejecta with $Y_\mathrm{e} \sim 0.47$ ($\approx 30/64$) predominantly during the QSE phase. On the other hand, for more massive models, its production of $\sim 40$--50\% is due to a $\nu$p-process (see a substantial enhancement at $A = 64$ with neutrinos in Fig.~\ref{fig:nuonoff}).

For low-mass models ($8.8\, M_\odot$ and $9.6\, M_\odot$), non-negligible production of a neutron-rich isotope $^{48}$Ca is found, whose origin remains a mystery. $^{48}$Ca, a double magic isotope with $(Z, N) = (20, 28)$, is produced in the neutron-rich ejecta with $Y_\mathrm{e} \sim 0.4$ ($\approx 20/48$) during the NSE and QSE phases (Fig.~\ref{fig:freeze}; left top). This implies that low-mass core-collapse supernovae can be, in part, possible sources of $^{48}$Ca \citep{Wanajo2013}, in addition to a hypothetical, rare class of thermonuclear supernovae from CO \citep{Meyer1996,Woosley1997} or ONe \citep{Jones2019,Jones2019b} cores. 

Neutrino-heated ejecta can also be sources of some radioactive isotopes such as $^{56}$Ni, $^{60}$Fe, and $^{44}$Ti. In particular, the mass of produced $^{56}$Ni serves as an important quantity for testifying theoretical models, which can be directly estimated by the observations of supernova light curves ($\sim 0.04\, M_\odot$ on average for Type~II supernovae, \citealt{Anderson2019}). $^{56}$Ni is synthesized through all stages of NSE, QSE, and an $\alpha$-process in only slightly neutron-rich or proton-rich conditions ($Y_\mathrm{e} > 0.49$). For the low-mass model ($8.8\, M_\odot$ or $9.6\, M_\odot$), the mass of $^{56}$Ni in the ejecta is $\sim 0.003\, M_\odot$ owing to its small mass of innermost ejecta ($\sim 10^{-2}\, M_\odot$) as well as the dominance of neutron-rich material ($Y_\mathrm{e} < 0.49$; top panels of Fig.~\ref{fig:yehist}), which is about one-tenth of a mean observational value. The small $^{56}$Ni mass is in fact consistent with the value estimated for some observed low-luminosity supernovae \citep{Hendry2005,Pastorello2007}, indicating such events originate from low-mass progenitors. For a more massive model ($15\, M_\odot$ or $27\, M_\odot$), the $^{56}$Ni mass is $\sim 0.006\, M_\odot$, which is noticeably smaller than the observational mean value, because the simulation was stopped while mass ejection was still continuing, as well as due to the omission of a part of the outer silicon layer. The recent hydrodynamical study of an exploding $19\, M_\odot$ star \citep{Bollig2021} shows that the production of $^{56}$Ni continues over seconds in the presence of long-lasting accretion flows, resulting in the total amount of $\sim 0.05\, M_\odot$ that is in agreement with the observational value.

$^{44}$Ti (half-life of 60~yr) is produced in the material with $Y_\mathrm{e} \sim 0.5$ mainly by an $\alpha$-process after the $\alpha$-rich freeze-out from QSE. In the models presented here, the mass of $^{44}$Ti in the ejecta is $\sim$~a few $10^{-6}\, M_\odot$, being substantially smaller than the inferred amounts of $\sim 1.0$--$1.7 \times 10^{-4}\, M_\odot$ for the supernova remnant Cassiopeia A \citep{Grefenstette2014,Siegert2015}. Similar to the case of $^{56}$Ni, the hydrodynamical simulation of a collapsing star including outer layers lasting over a long period of time is required for a quantitative prediction of $^{44}$Ti production. A recent study of nucleosynthesis based on the long-term core-collapse simulation of a $15\, M_\odot$ progenitor shows the production of $^{44}$Ti with a similar amount to the observational value \citep{Wongwathanarat2017}. 

For the low-mass model ($8.8\, M_\odot$ or $9.6\, M_\odot$), $^{60}$Fe (half-life of 2.62~Myr) is produced under neutron-rich conditions ($Y_\mathrm{e} \approx 0.42$--$0.43 \sim 26/60$) in NSE and subsequent $\alpha$-deficient QSE \citep{Wanajo2013b}. Such low-$Y_\mathrm{e}$ material is absent for more massive models ($15\, M_\odot$ and $27\, M_\odot$), resulting in little production of $^{60}$Fe in their neutrino-heated ejecta. The mass of synthesized $^{60}$Fe in the low-mass model is $\sim 3$--$4\times 10^{-5}\, M_\odot$, which is comparable to the estimated amount produced in the outer envelope (that is not included in the analysis presented here) by successive neutron captures on Fe isotopes \citep{Sukhbold2016}. This implies that a part of live $^{60}$Fe in the Milky Way \citep{Smith2004,Harris2005} has been made in the neutrino-heated ejecta from low-mass core-collapse supernovae \citep{Wanajo2013b}.

\section{\textit{Nucleosynthesis in neutrino-driven winds}}
\label{sec:winds}

About one or a few seconds after core bounce, the dense material has been evacuated from the surface region of a proto-neutron star in the form of subsonic neutrino-heated ejecta, leaving behind a low density region. Subsequent ejecta from the surface of the proto-neutron star become transonic outflows, known as neutrino-driven winds, lasting over 10 seconds. Under such conditions, the properties of neutrino-driven outflows can be well described as the spherically symmetric, stationary transonic solutions of a suite of relevant equations described below \citep{Duncan1986,Qian1996,Cardall1997,Otsuki2000,Wanajo2001,Thompson2001}. In fact, long-term one-dimensional core-collapse simulations over $\sim 10$ seconds \citep{Hudepohl2010,Fischer2012,Roberts2012} indicate overall similar physical properties of neutrino-driven winds to those obtained by such stationary wind models. Because of demanding computational costs, there have been few multi-dimensional simulations of neutrino-driven winds, although the effects of anisotropy are expected to be subdominant unless rotational velocity or magnetic field of a proto-neutron star is fairly large. It is important to note that, in the presence of long-lasting (for a duration of several seconds) accretion flows toward a proto-neutron star, neutrino-driven winds can be absent or emerge only when neutrino emission becomes weak as indicated by recent three-dimensional work \citep{Mueller2017,Bollig2021}. Thus, neutrino-driven winds may appear only in the case of a relatively low-mass progenitor that contains a small amount of dense outer material \citep{Stockinger2020}.

\subsection{\textit{Stationary wind solutions}}
\label{subsec:solution}

The spherically symmetric, general-relativistic models of neutrino-driven winds described here are those developed in \citet{Otsuki2000,Wanajo2001}. Since the ejected mass is assumed to be negligible compared to the mass of the proto-neutron mass $M$, the gravitational field in which neutrino-heated material moves can be treated as a fixed background Schwarzschild metric. The equations of mass, momentum, and energy conservation are written as
\begin{equation}
    \dot{M} = 4 \pi r^2 \rho u,
    \label{eq:mass}
\end{equation}
\begin{equation}
    u \frac{du}{dr} = -\frac{1 + (u/c)^2 - 2GM/c^2r}{\rho (1 + \epsilon/c^2) + P/c^2} \frac{dP}{dr} - \frac{GM}{r^2},
    \label{eq:momentum}
\end{equation}
\begin{equation}
    \dot{q} = u \left(\frac{d\epsilon}{dr} - \frac{P}{\rho^2} \frac{d\rho}{dr}\right),
    \label{eq:energy}
\end{equation}
where $\dot{M}$ is the mass ejection rate, $\dot{q}$ is the net heating rate, $\rho$ is the mass density, $P$ is the pressure,  $\epsilon$ is the specific internal energy, $r$ is the distance from the center of the neutron star, and $G$ is the gravitational constant. Note that the effects of general relativity are crucial for describing the properties of neutrino-driven winds \citep{Cardall1997,Otsuki2000}. The velocity $u$ here is related to the proper velocity $v$ of the matter measured by a local, stationary observer by
\begin{equation}
    v = \left[1 + \left(\frac{u}{c}\right)^2 - \frac{2GM}{c^2r}\right]^{-1/2} u.
\end{equation}

The net heating rate $\dot{q}$ is given by
\begin{equation}
    \dot{q} \approx \dot{q}_{\nu N} + \dot{q}_{\nu e} + \dot{q}_{\nu \nu} - \dot{q}_{eN} - \dot{q}_{ee},
    \label{eq:net}
\end{equation}
where heating is due to electron neutrino and electron anti-neutrino capture on free nucleons in Eqs.~(\ref{eq:nucap}) and (\ref{eq:anucap}), $\dot{q}_{\nu N}$, neutrino scattering by electrons and positrons, $\dot{q}_{\nu e}$, and neutrino-antineutrino pair annihilation into electron-positron pairs, $\dot{q}_{\nu \nu}$. Cooling is due to electron and positron capture on free nucleons (inverse of Eqs.~(\ref{eq:nucap}) and (\ref{eq:anucap})), $\dot{q}_{eN}$, and electron-positron pair annihilation into neutrino-antineutrino pairs, $\dot{q}_{ee}$. These components (in units of MeV/g/s) can be computed by \citep{Otsuki2000}
\begin{equation}
    \dot{q}_{\nu N} \approx \frac{9.65}{m_\mathrm{u}} \left[(1 - Y_\mathrm{e}) L_{\nu_\mathrm{e},51} \frac{\langle E_{\nu_\mathrm{e}}^3\rangle}{\langle E_{\nu_\mathrm{e}}\rangle} + Y_\mathrm{e} L_{\bar{\nu}_\mathrm{e},51} \frac{\langle E_{\bar{\nu}_\mathrm{e}}^3\rangle}{\langle E_{\bar{\nu}_\mathrm{e}}\rangle}\right] \frac{1 - g_1(r)}{R_{\nu 6}^2} \Phi(r)^6,
    \label{eq:nun}
\end{equation}
\begin{equation}
    \dot{q}_{\nu e} \approx \frac{2.17}{m_\mathrm{u}} \frac{T_\mathrm{MeV}^4}{\rho_8} \left(L_{\nu_\mathrm{e},51} \epsilon_{\nu_\mathrm{e}} + L_{\bar{\nu}_\mathrm{e},51} \epsilon_{\bar{\nu}_\mathrm{e}} + \frac{6}{7} L_{\nu_\mathrm{\mu},51} \epsilon_{\nu_\mathrm{\mu}}\right) \frac{1 - g_1(r)}{R_{\nu 6}^2} \Phi(r)^5,
    \label{eq:nue}
\end{equation}
\begin{equation}
    \dot{q}_{\nu\nu} \approx \frac{12.0}{m_\mathrm{u}} \left[L_{\nu_\mathrm{e},51} L_{\bar{\nu}_\mathrm{e},51} (\epsilon_{\nu_\mathrm{e}} + \epsilon_{\bar{\nu}_\mathrm{e}}) + \frac{6}{7} L_{\nu_\mathrm{\mu},51}^2 \epsilon_{\nu_\mathrm{\mu}}\right] \frac{g_2(r)}{\rho_8 R_{\nu 6}^4} \Phi(r)^9,
    \label{eq:nunu}
\end{equation}
\begin{equation}
    \dot{q}_{eN} \approx \frac{2.27}{m_\mathrm{u}} T_\mathrm{MeV}^6,
    \label{eq:en}
\end{equation}
and
\begin{equation}
    \dot{q}_{ee} \approx \frac{0.144}{m_\mathrm{u}} \frac{T_\mathrm{MeV}^9}{\rho_8},
    \label{eq:ee}
\end{equation}
where $1-g_1(r)$ is a gravitational geometrical factor that represents the effect of bending a neutrino trajectory with $g_1(r)$ given by
\begin{equation}
    g_1(r) = \left[1 - \left(\frac{R_\nu}{r}\right)^2 \frac{1 - 2GM/c^2r}{1 - 2GM/c^2R_\nu}\right]^{1/2},
    \label{eq:bend}
\end{equation}
while $g_2(r)$ is given by
\begin{equation}
    g_2(r) = \left[1 - g_1(r)\right]^4 \left[g_1(r)^2 + 4g_1(r) + 5\right],
    \label{eq:bend2}
\end{equation}
and $\Phi(r)$ is the gravitational redshift factor defined by
\begin{equation}
    \Phi(r) = \left(\frac{1 - 2GM/c^2R_\nu}{1 - 2GM/c^2r}\right)^{1/2}.
    \label{eq:redshift}
\end{equation}
Here, $R_\nu$ is the radius of the neutrinosphere, $R_{\nu 6}$ is $R_\nu$ in units of $10^6$~cm, $\rho_8$ is the matter density in units of $10^8$~g/cm$^3$, $T_\mathrm{MeV}$ is the temperature in units of MeV, $\langle E_{\nu_\mathrm{e}}^n\rangle$ is the $n$th energy moment of the electron neutrino energy distribution in units of MeV$^n$, $\epsilon_{\nu_\mathrm{e}} \equiv \langle E_{\nu_\mathrm{e}}^2\rangle/\langle E_{\nu_\mathrm{e}}\rangle$ is the mean energy of an electron neutrino in units of MeV, and $L_{\nu_\mathrm{e},51}$ is the luminosity of electron neutrinos in units of $10^{51}$~erg/s (same for other flavors). Note that all heating terms are dependent on general-relativistic effects, i.e., the geometrical and redshift factors. The former effect enhances the heating rate, while the latter effect reduces it. According to \citet{Otsuki2000}, the net effects of these factors work to substantially increase the heating rate compared to those in the Newtonian gravity. As can be seen in Eqs.~(\ref{eq:bend})--(\ref{eq:redshift}), general-relativistic effects appear with the term $2GM/c^2R_\nu \propto M/R_\nu$, the reciprocal of the neutrinosphere radius with respect to the Schwarzschild radius, which indicates the compactness of the proto-neutron star. As a result, the heating rates increase for more massive or smaller proto-neutron stars. 

Hereafter, several assumptions are made: $R_\nu$ is the same as the proto-neutron star radius, $R$, the neutrino luminosity of a single flavor, $L_\nu$, is the same to each other (i.e., the total luminosity is $6L_\nu$), $\langle E_{\nu_\mathrm{e}}^3\rangle/\langle E_{\nu_\mathrm{e}}\rangle = 1.14 \epsilon_{\nu_\mathrm{e}}$ (same for other flavors) and the mean neutrino energies of 12, 14, and 14~MeV for electron neutrino, electron antineutrino, and heavy lepton neutrinos, respectively (according to, e.g., \citealt{Qian1996,Janka2012}). For a given combination of $M$, $R$, and $L_\nu$, the suite of ordinary differential equations, Eqs.~(\ref{eq:mass})--(\ref{eq:energy}), can be numerically solved (see, e.g., \citealt{Thompson2001}) with the equations of state for ions and arbitrary relativistic or degenerate electrons and positrons (e.g., \citealt{Timmes2000}). The boundary conditions can be chosen to be, e.g., $\rho(R) = 10^{10}$~g/cm$^3$, the temperature $T(R)$ is determined by the condition $\dot{q}(R) = 0$ (at which heating and cooling balance), and $u(R)$ such that the outflow becomes transonic.

\begin{figure}
    \centering
    \includegraphics[width=0.49\textwidth]{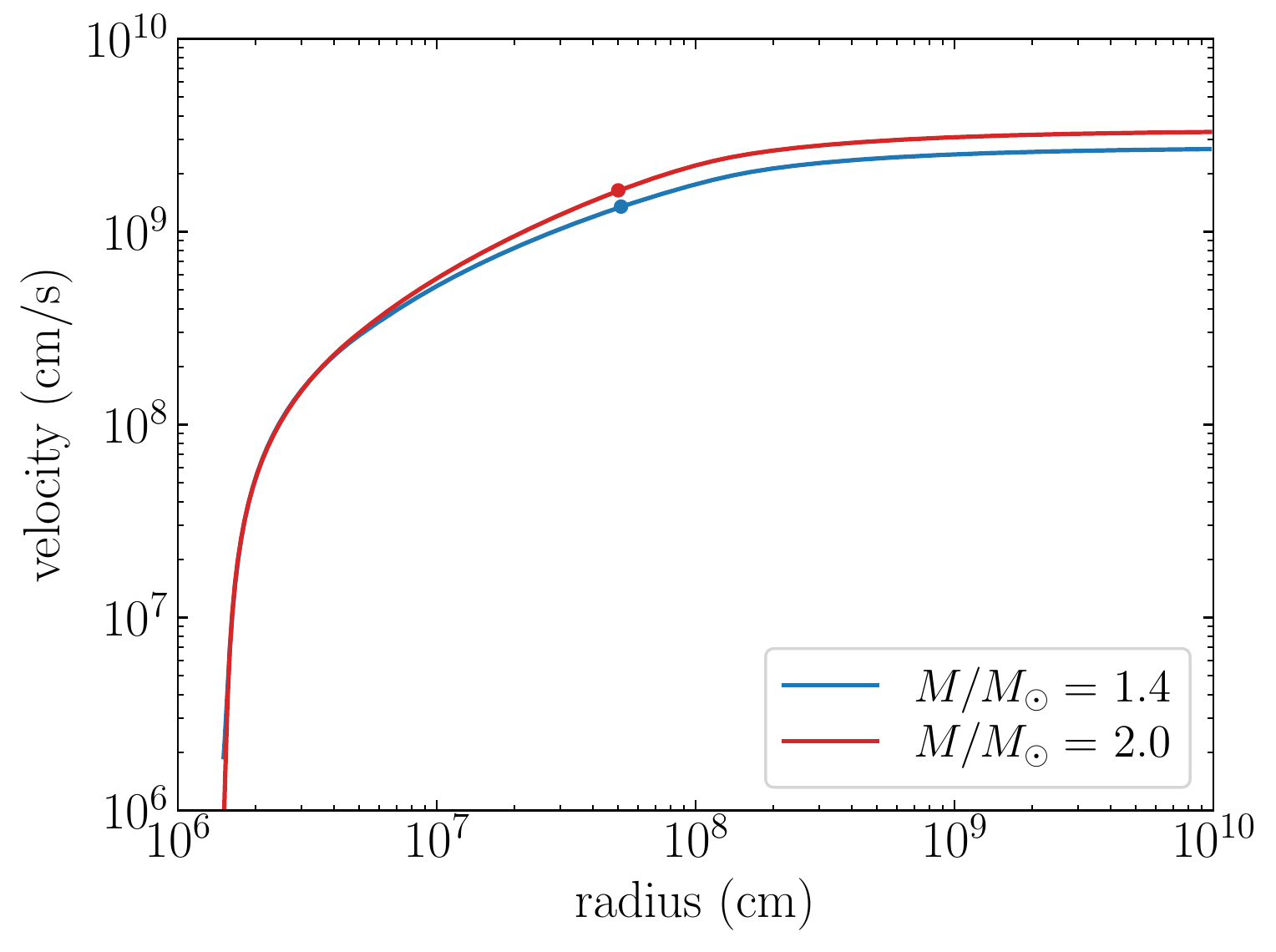}
    \includegraphics[width=0.49\textwidth]{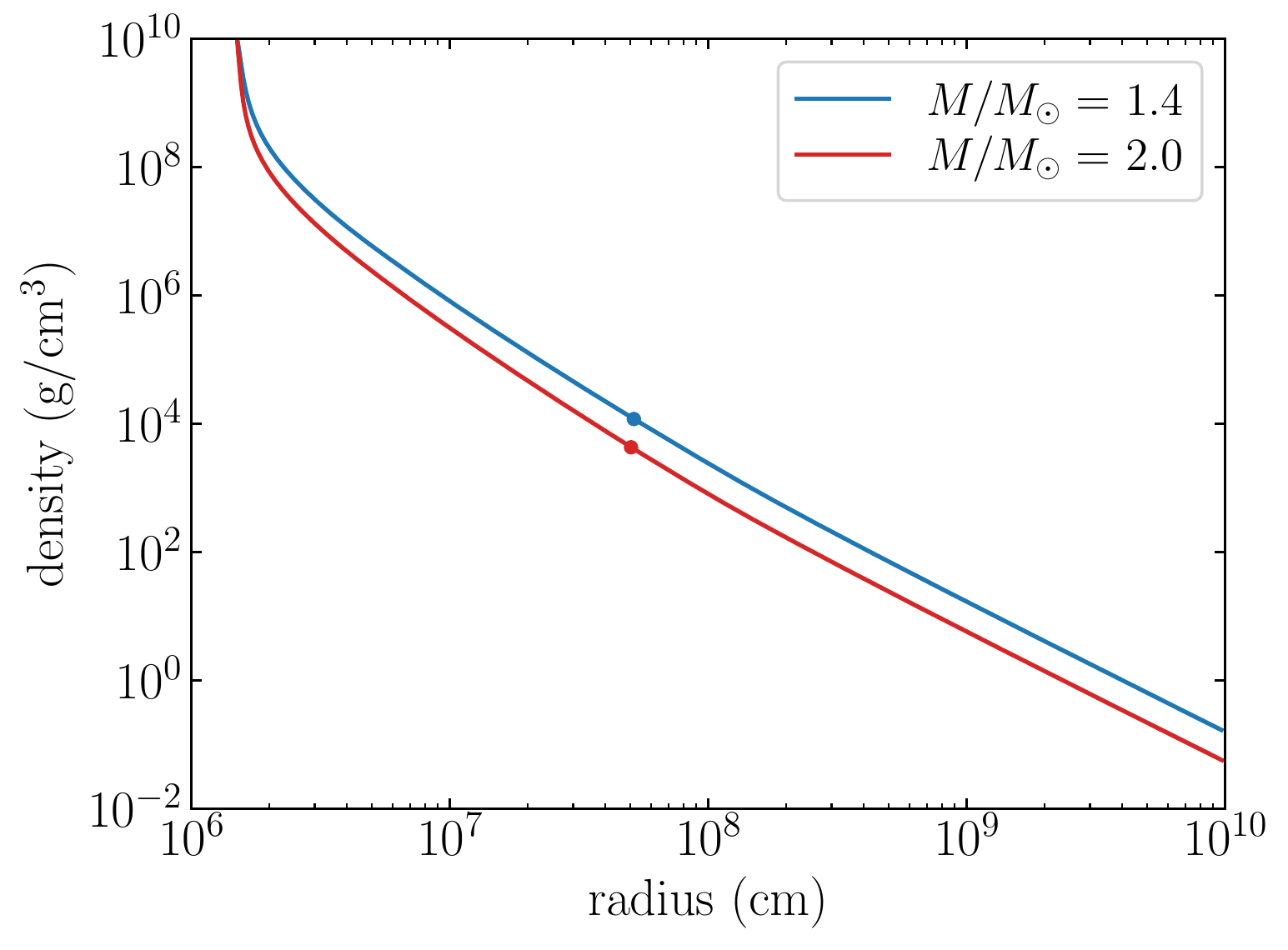}
    \includegraphics[width=0.49\textwidth]{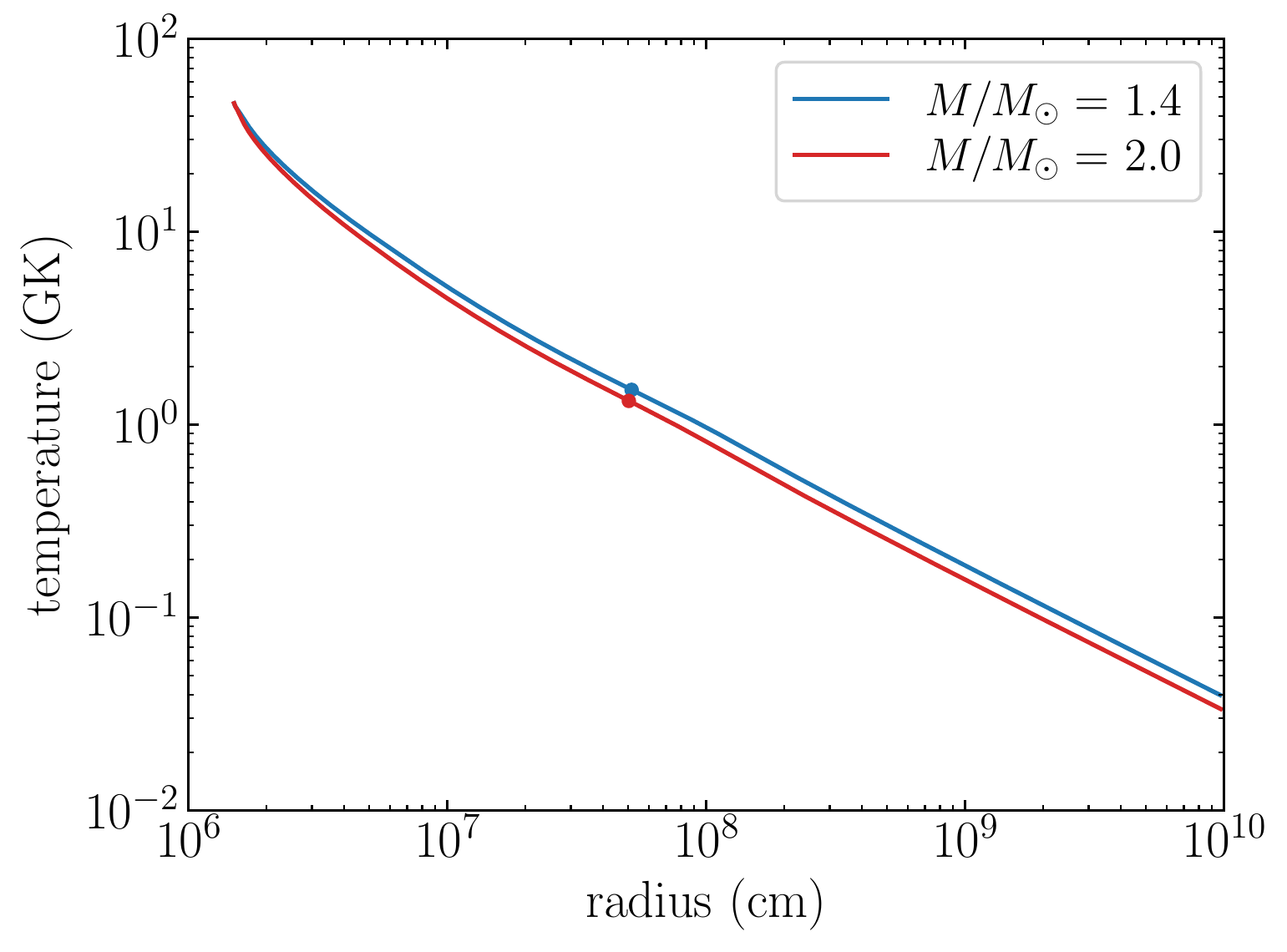}
    \includegraphics[width=0.49\textwidth]{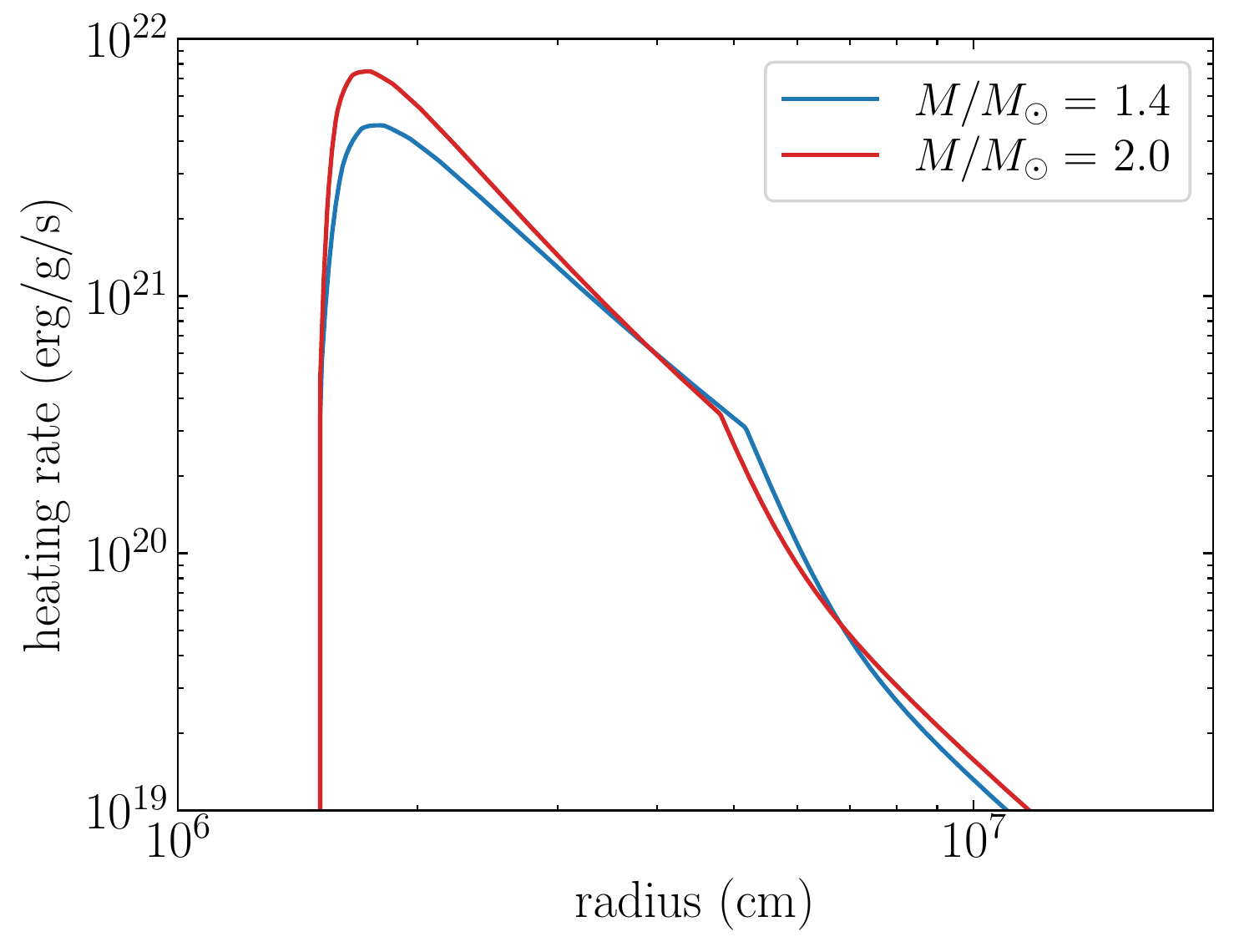}
    \caption{Properties of neutrino-driven winds for a proto-neutron star with neutrino luminosity of $L_\nu = 10^{52}$~erg/s, radius of $R = 15$~km, and mass of $M = 1.4\, M_\odot$ (blue) or $2.0\, M_\odot$ (red). The top-left, top-right, and bottom-left panels display, respectively, the velocity ($u$), density, and temperature of outgoing material as functions of radius. On each line, the circle marks the sonic radius (i.e., the sonic velocity in the top-left panel). The bottom-right panel shows the heating rates as functions of radius. A sudden drop of the rate at $\sim 50$~km is due to recombination of free nucleons to $\alpha$ particles.}
    \label{fig:wind}
\end{figure}

Fig.~\ref{fig:wind} displays the solutions of neutrino-driven winds as functions of radius ($u$, $\rho$, and $T$ in left-top, right-top, and left-bottom panels, respectively) for $L_\nu = 10^{52}$~erg/s, $R = 15$~km, and $M = 1.4\, M_\odot$ (blue) or $2.0\, M_\odot$ (red). As can be seen in the top-left panel, an initially subsonic outflow becomes supersonic (i.e. transonic) through the sonic point (circle). Note that an initial velocity below and above $u(R)$ adopted here gives subsonic (i.e. the outflow eventually returns to the proto-neutron star surface) and unphysical solutions, respectively. The bottom-right panel shows the net heating rates as functions of radius. It is found that neutrino heating plays a role only in the vicinity of the proto-neutron star surface, $r \le 2 R$, which becomes unimportant at a larger radius (in part owing to nucleon recombination to $\alpha$ particles). For a fixed $R$, a more massive case ($M = 2.0\, M_\odot$) results in a higher heating rate and a faster outflow predominantly due to its greater compactness $M/R$ (the same holds true for a fixed $M$ and a smaller $R$). 

\begin{figure}
    \centering
    \includegraphics[width=0.6\textwidth]{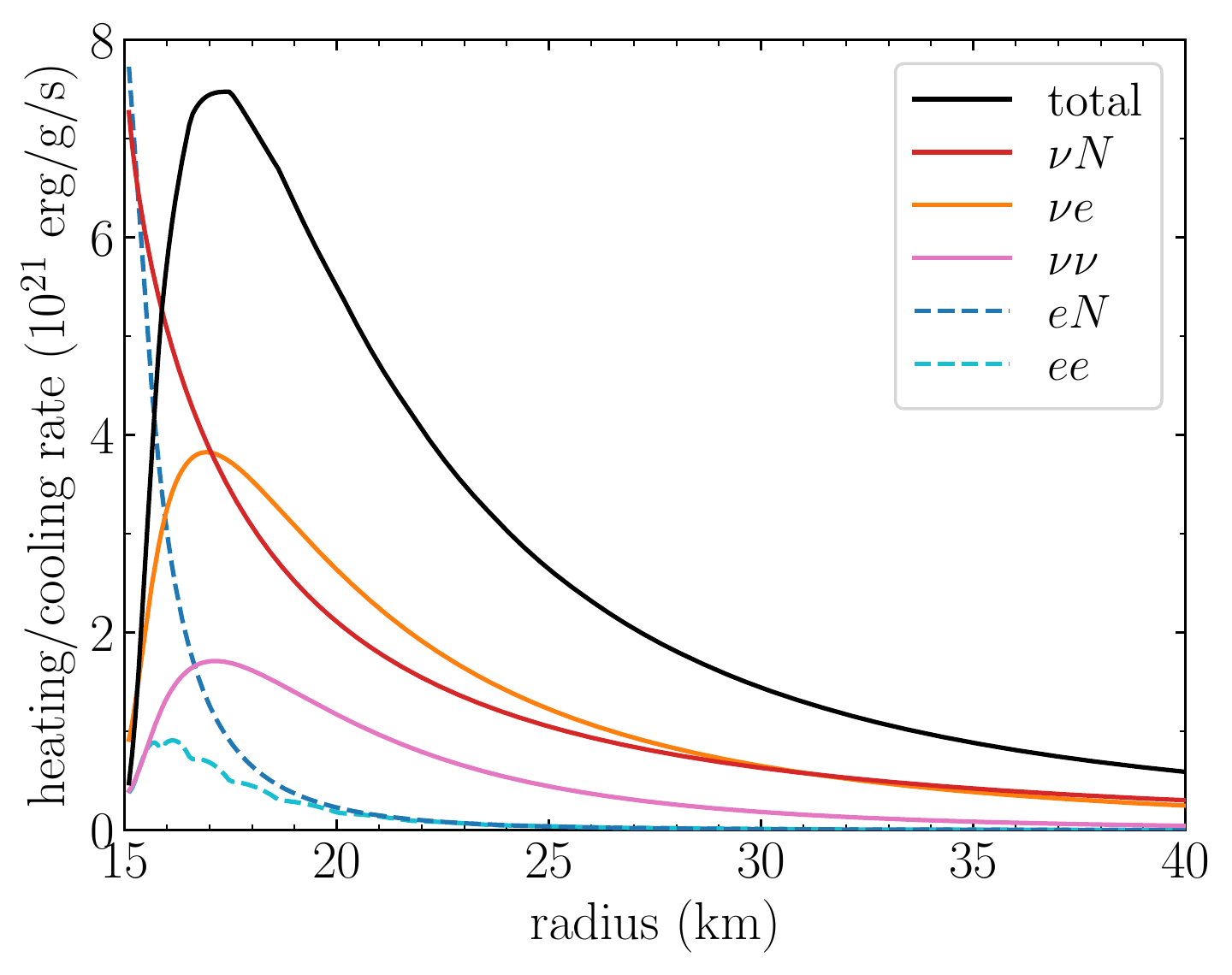}
    \caption{Net heating rate (black curve) and its decomposition into heating (solid color curves) and cooling (dashed curves) terms for a proto-neutron star with neutrino luminosity of $10^{52}$~erg/s, radius of 15~km, and mass of $2.0\, M_\odot$. Heating is due to neutrino capture on free nucleons ($\nu N$), neutrino scattering by electrons and positrons ($\nu e$), and neutrino-antineutrino pair annihilation into electron-positron pairs ($\nu\nu$). Cooling is due to electron and positron capture on free nucleons ($eN$) and electron-positron pair annihilation into neutrino-antineutrino pairs ($ee$).}
    \label{fig:qdot}
\end{figure}

The contribution of each component in Eq.~(\ref{eq:net}) to the net heating rate is shown in Fig.~\ref{fig:qdot} for $L_\nu = 10^{52}$~erg/s, $R = 15$~km, and $M = 2.0\, M_\odot$. For heating terms (solid, colored curves), dominant contributors are neutrino capture on free nucleons ($\nu N$) and neutrino scattering by electrons and positrons ($\nu e$), while neutrino-antineutrino pair annihilation into electron-positron pairs ($\nu\nu$) plays a subdominant role (but see its substantial contribution in case of highly non-spherical configurations, \citealt{Wanajo2006b,Wanajo2012}). Compared to heating terms, cooling terms (dashed curves) rapidly decay with increasing radius (and thus decreasing temperature) because of their high sensitivities to temperature ($\dot{q}_{eN} \propto T^6$ for electron and positron capture on free nucleons and $\dot{q}_{ee} \propto T^9/\rho$ for electron-positron annihilation into neutrino-antineutrino pairs).

\subsection{\textit{(No) r-process in neutrino-driven winds}}
\label{subsec:rpro}

The neutron-richness, or $Y_\mathrm{e}$, in neutrino-driven ejecta is predominantly determined by the equilibrium condition for neutrino capture on free nucleons in Eqs.~(\ref{eq:nucap}) and (\ref{eq:anucap}) such as \citep{Qian1996}
\begin{equation}
    Y_\mathrm{e,eq} \approx \left(1 + \frac{L_{\bar{\nu}_\mathrm{e}}}{L_{\nu_\mathrm{e}}} \frac{\epsilon_{\bar{\nu}_\mathrm{e}}-2\Delta+1.2\Delta^2/\epsilon_{\bar{\nu}_\mathrm{e}}}{\epsilon_{\nu_\mathrm{e}}+2\Delta+1.2\Delta^2/\epsilon_{\nu_\mathrm{e}}}\right)^{-1},
    \label{eq:yeeq}
\end{equation}
where $\Delta = 1.293$~MeV is the neutron-proton mass difference. Assuming $L_{\nu_\mathrm{e}} \approx L_{\bar{\nu}_\mathrm{e}}$ and omitting the terms of $\Delta^2$, the neutron-rich condition of $Y_\mathrm{e,eq} < 0.5$ becomes
\begin{equation}
    \epsilon_{\bar{\nu}_\mathrm{e}} - \epsilon_{\nu_\mathrm{e}} > 4\Delta \sim 5\, \mathrm{MeV}.
    \label{eq:ediff}
\end{equation}
However, recent hydrodynamical studies show $\epsilon_{\bar{\nu}_\mathrm{e}} - \epsilon_{\nu_\mathrm{e}} \sim 2$~MeV, a value too small to achieve neutron-rich conditions in neutrino-driven winds. In fact, long-term core-collapse simulations indicate that the ejecta can be only slightly neutron-rich ($Y_\mathrm{e} \sim 0.4$--0.5) at early times and become proton-rich ($Y_\mathrm{e} > 0.5$) at late times \citep{Hudepohl2010,Fischer2012,Roberts2012}. This subsection explores the requisite conditions such that an r-process occurs in such slightly neutron-rich neutrino-driven winds, mainly based on \citet{Wanajo2013c}.


The physical conditions needed for a successful r-process can be described by entropy $S$ ($\propto T^3/\rho$, assuming contributions from radiation and relativistic electron-positron pairs) and expansion timescale $t_\mathrm{exp}$ (defined as the duration from $T_\mathrm{GK} = 9$ to 2.5) in addition to $Y_\mathrm{e}$ \citep{Hoffman1996}. An r-process begins as the temperature decreases to $T_\mathrm{GK} \sim 2.5$, at which the nuclei of $A \sim 80$--90 formed in QSE act as ``seeds" that capture free nucleons (e.g., Fig.~\ref{fig:freeze}, top panels). That is, the number of free neutrons with respect to that of heavy nuclei, $Y_\mathrm{n}/Y_\mathrm{heavy}$, must be greater than $\sim 100$, so that nuclear flow can reach heavy r-process nuclei of $A \sim 200$. In early neutrino-heated ejecta, the values of $Y_\mathrm{n}/Y_\mathrm{heavy}$ are below unity at the end of the QSE phase ($T_\mathrm{GK} \sim 4$; right panel of Fig.~\ref{fig:toseed}), resulting in no r-process. However, higher values of $Y_\mathrm{n}/Y_\mathrm{heavy}$ can be expected in neutrino-driven winds owing to their higher $S$ and shorter $t_\mathrm{exp}$. A higher $S$, i.e., a lower $\rho$ at a given $T$, reduces the rate of the forward reaction (the gateway to heavy nuclei; being proportional to $\rho^2$) in Eq.~(\ref{eq:aan}), resulting a higher $Y_\mathrm{n}/Y_\mathrm{heavy}$ at the beginning of an r-process. A short $t_\mathrm{exp}$ also plays a role in maintaining a high $Y_\mathrm{n}/Y_\mathrm{heavy}$ because of the insufficient period of time for the production of heavy nuclei through the three-body reaction in Eq.~(\ref{eq:aan}).

\begin{figure}
    \centering
    \includegraphics[width=1\textwidth]{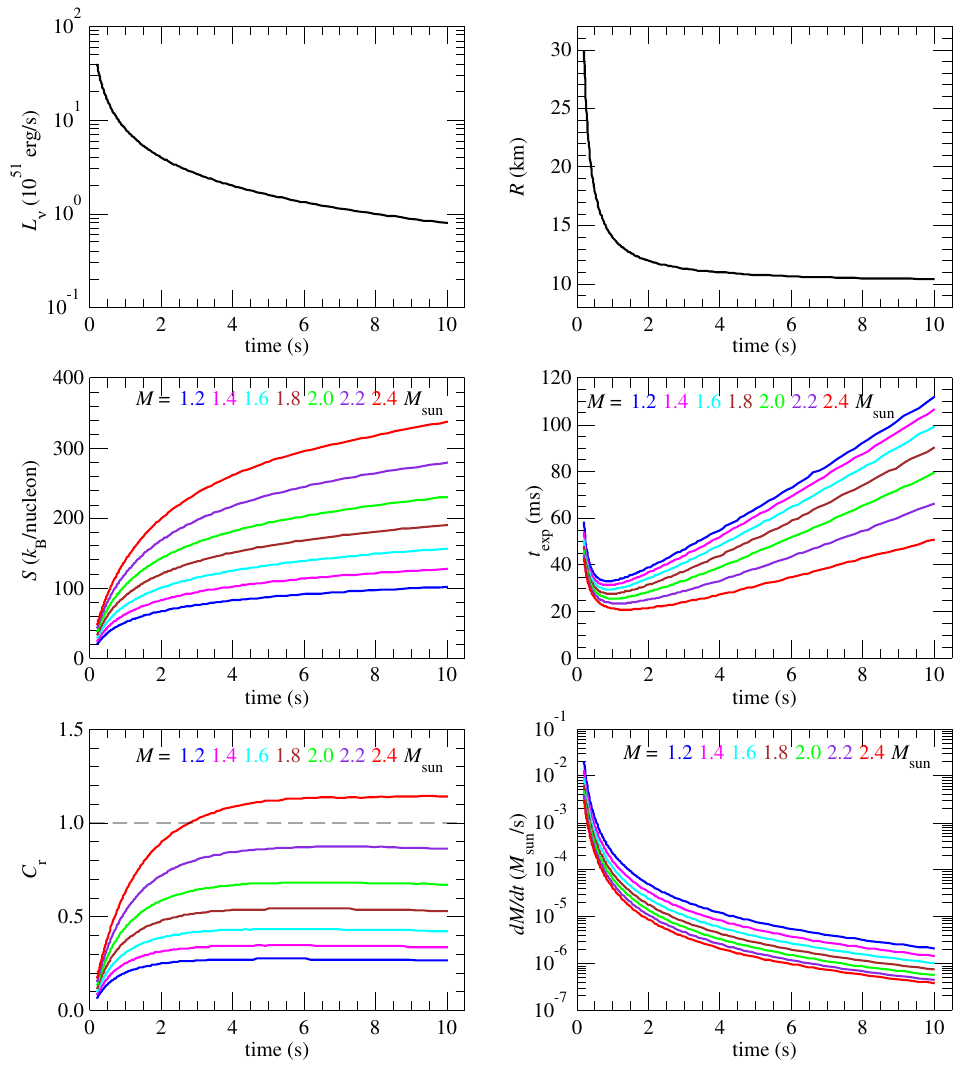}
    \caption{Temporal evolution of entropy ($S$, middle-left panel), expansion timescale ($t_\mathrm{exp}$, middle-right), the criterion for a successful r-process ($C_\mathrm{r}$ in Eq.~(\ref{eq:cr} for $Y_\mathrm{e} = 0.4$), bottom-left), and mass ejection rate ($\dot{M}$ in Eq.~(\ref{eq:mass}); bottom-right) for various proto-neutron star masses ($M/M_\odot = 1.2,\, \cdots, 2.4$), assuming the neutrino luminosity ($L_\nu$) and radius ($R$) shown in the top-left and top-right panels, respectively (see text). The dashed line in the bottom-left panel marks $C_\mathrm{r} = 1$, above which r-process nuclei with $A \sim 200$ are expected to be appreciably produced. (\citealt{Wanajo2013c}; reproduced with permission of the AAS.)}
    \label{fig:property}
\end{figure}

According to \citet{Hoffman1996}, the condition for producing the third-peak r-process abundances ($A\sim 200$) can be expressed by these three quantities,  $Y_\mathrm{e}$, $S$, and $t_\mathrm{exp}$, as
\begin{equation}
    C_\mathrm{r} \equiv 0.0005\, \frac{S}{{Y_\mathrm{e}\, t_\mathrm{exp}}^{1/3}} > 1
    \label{eq:cr}
\end{equation}
for $0.4 < Y_\mathrm{e} < 0.5$ (slightly neutron-rich condition relevant to neutrino-driven winds; see \citealt{Hoffman1996,Fujibayashi2020} for $Y_\mathrm{e} < 0.4$). As can be found in Eq.~(\ref{eq:cr}), a combination of high $S$, short $t_\mathrm{exp}$, and low $Y_\mathrm{e}$ is favorable for a successful r-process. 

\begin{figure}
    \centering
    \includegraphics[width=0.6\textwidth]{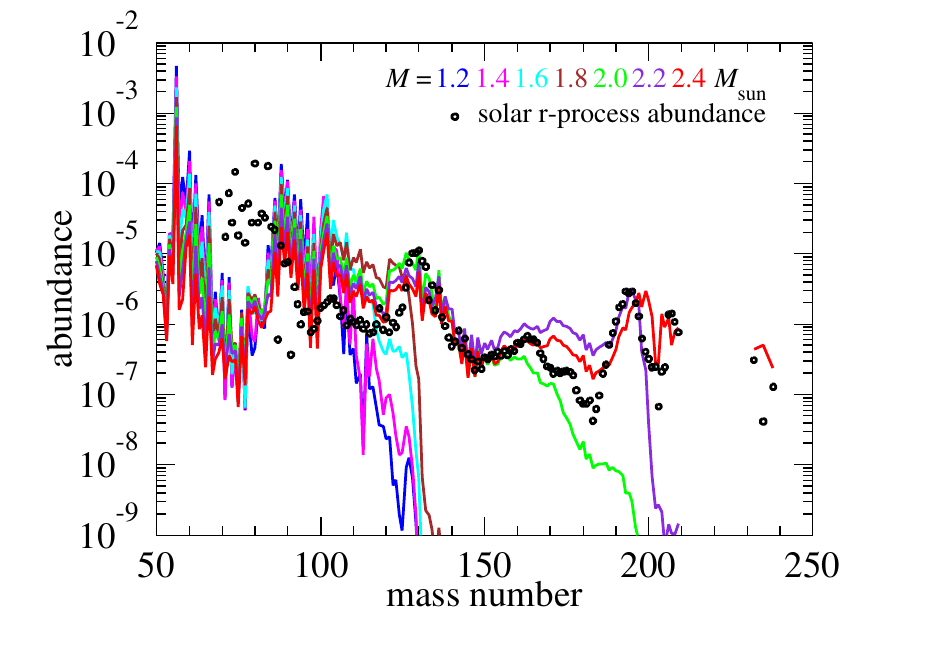}
    \caption{Final abundances for various proto-neutron star masses ($M/M_\odot = 1.2,\, \cdots, 2.4$), assuming the temporal evolution of neutrino luminosity ($L_\nu$) and radius ($R$) shown in top-left and top-right panels of Fig.~\ref{fig:property}, respectively. The circles show the solar r-process residuals \citep{Prantzos2020} vertically shifted to match the height of the third peak ($A \sim 200$) for $M/M_\odot = 2.4$. (\citealt{Wanajo2013c}; reproduced with permission of the AAS.)}
    \label{fig:ymav}
\end{figure}

Fig.~\ref{fig:property} shows the temporal evolution of $S$ (middle-left panel), $t_\mathrm{exp}$ (middle-right), $C_\mathrm{r}$ with $Y_\mathrm{e} = 0.4$ (defined by Eq.~(\ref{eq:cr}); bottom-left), and $\dot{M}$ (in Eq.~(\ref{eq:mass}; bottom-right) as a series of time-stationary wind solutions. The use of time-stationary solutions is reasonable due to fairly shorter dynamical timescales than the evolutionary time of $L_\nu$ and $R$. Here, the temporal evolution of these quantities is assumed to be $L_\nu(t) = L_{\nu,0} (t/t_0)^{-1}$ with $L_{\nu,0} = 4\times 10^{52}$~erg/s and $t_0 = 0.2$~s (left top) and $R(t) = (R_0 - R_1) (L_\nu/L_{\nu,0}) + R_1$ with $R_0 = 30$~km and $R_1 = 10$~km (right top) such that each wind solution can be obtained from a given set of $(M, L_\nu)$. A wide range of proto-neutron mass is considered, ranging from $M =1.2\, M_\odot$ (nearly a minimum estimated value, \citealt{Sukhbold2016}) to $2.4\, M_\odot$ (a value close to the causality limit of $R > 4.3\, (M/M_\odot)$~km with $R \sim 10$~km, \citealt{Lattimer2011}, i.e., the speed of sound must not exceed the speed of light).

For a given $M$, $S$ and $t_\mathrm{exp}$ increase with time (after $t \sim 1$~s at which $R$ becomes sufficiently small). Because of the greater contribution of $S$ in Eq.~(\ref{eq:cr}), the value of $C_\mathrm{r}$ also increases with time. Moreover, $S$ and $t_\mathrm{exp}$ are larger and smaller, respectively, for a more massive proto-neutron star. As a result, the value of $C_\mathrm{r}$ at a given time is greater for a higher $M$ as can be seen in the bottom-left panel of Fig.~\ref{fig:property}. Nevertheless, the value of $C_\mathrm{r}$ exceeds unity only for $M = 2.4\, M_\odot$ (and marginally for $2.2\, M_\odot$). In practice, the high-mass end of proto-neutron stars is estimated to be $\sim 1.6\, M_\odot$ \citep{Sukhbold2016}, with which the value of $C_\mathrm{r}$ reaches no more than $\sim 0.5$. Therefore, it is concluded that neutrino-driven winds from core-collapse supernovae are unlikely sources of heavy r-process elements.

Nucleosynthetic abundances are shown in Fig.~\ref{fig:ymav}, which compares these with the solar r-process residuals (i.e., compared to the solar abundances, but subtracting the s-process component, \citealt{Prantzos2020}). Here, the temporal evolution of $Y_\mathrm{e}$ is assumed to take the minimum value of 0.4 at $3$\,s (see \citealt{Wanajo2013}). As anticipated from the above discussion, only the case of $M = 2.4\, M_\odot$ (and marginally of $2.2\, M_\odot$) results in the production of heavy r-process nuclei up to actinides (Th and U).
However, for a plausible range of proto-neutron star masses, $M/M_\odot = 1.2$--1.6 \citep{Sukhbold2016}, only light r-process nuclei with $A \sim 90$--100 are synthesized. Therefore, it is likely that neutrino-driven winds can be sources of only some lightest r-process elements from Sr to Mo.

\subsection{\textit{$\nu$p-process in neutrino-driven winds}}
\label{subsec:nuppro}

As noted in the previous subsection, recent long-term core-collapse simulations indicate that neutrino-driven winds become proton-rich \citep{Hudepohl2010,Fischer2012,Roberts2012} because of the small energy differences in electron neutrinos and electron anti-neutrinos (as oppose to the condition in Eq.~(\ref{eq:ediff})). Therefore, a $\nu$p-process \citep{Froehlich2006,Pruet2006,Wanajo2006} is expected to play a role for the production of proton-rich heavy nuclei in neutrino-driven winds. This subsection provides several aspects of nucleosynthetic outcomes in proton-rich environments based on the results in \citet{Wanajo2011b} with the use of time-stationary wind solutions described in previous subsections.

\begin{figure}
    \centering
    \includegraphics[width=0.6\textwidth]{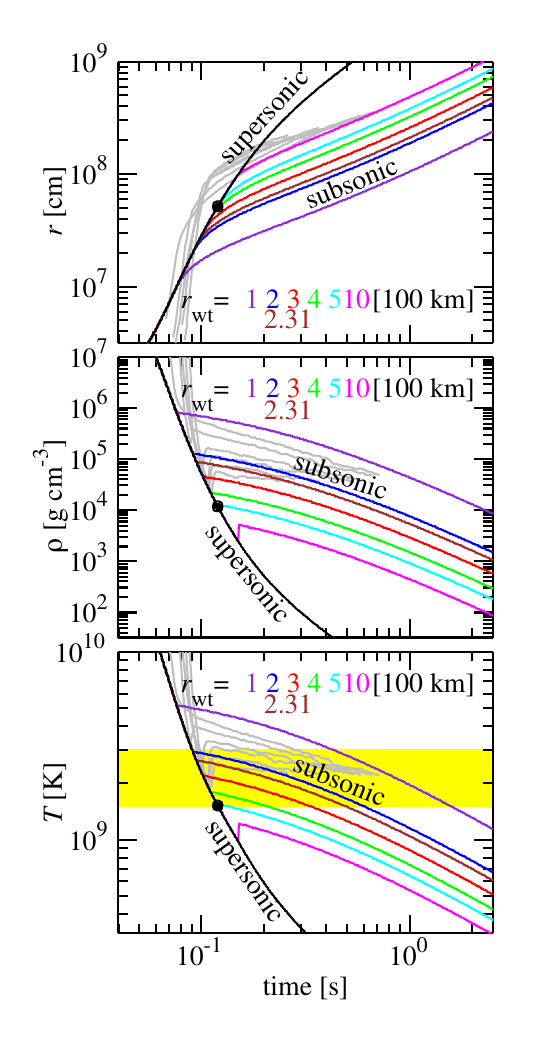}
    \caption{Radius (top), density (middle), and temperature (bottom) as a function of time for the wind solution with $L_\nu = 10^{52}$~erg/s and $M = 1.4\, M_\odot$. The curves with different colors show the subsonic outflows after wind termination at $r_\mathrm{wt} = 100$, 200, 231, 300, 400, 500, and 1000~km. For $r_\mathrm{wt} = 1000$~km, the Rankine-Hugoniot shock-jump conditions are applied at wind termination. The black curve shows the supersonic outflow without wind termination. The filled circle in each panel marks the sonic point. The yellow band in the bottom panel indicates the temperature range ($T_\mathrm{GK} = 1.5$--3) relevant to the $\nu$p-process. The wind trajectories from hydrodynamical results by \cite{Buras2006} are shown by gray curves for comparison purposes. (\citealt{Wanajo2011b}; reproduced with permission of the AAS.)}
    \label{fig:tjt}
\end{figure}

First, the solution for $L_\nu = 10^{52}$~erg/s and $M = 1.4\, M_\odot$ (and $R = 15$~km according to the top panels in Fig.~\ref{fig:property}) is considered as representative of early neutrino-driven winds (where neutrino irradiation is sufficiently strong) as shown in Fig.~\ref{fig:tjt}. Here, wind termination by slowly outgoing, early neutrino-heated ejecta is assumed at the various radii of $r_\mathrm{wt} = 100$, 200, 231, 300, 400, 500, and 1000~km. The density after wind termination is assumed to follow $\rho \propto t^{-2}$ (and thus $T \propto t^{-2/3}$ with $S \propto T^3/\rho =$ const.) according to hydrodynamical results (e.g., \citealt{Arcones2007}). The evolution of radius can be obtained from $\dot{M} = 0$ in Eq.~(\ref{eq:mass}). The yellow band in the bottom panel indicates the temperature range ($T_\mathrm{GK} = 1.5$--3) relevant to the $\nu$p-process, this being sufficiently lower than the freezeout temperature from QSE ($T_\mathrm{GK} \sim 4$), but high enough for proton capture.

\begin{figure}
    \centering
    \includegraphics[width=0.46\textwidth]{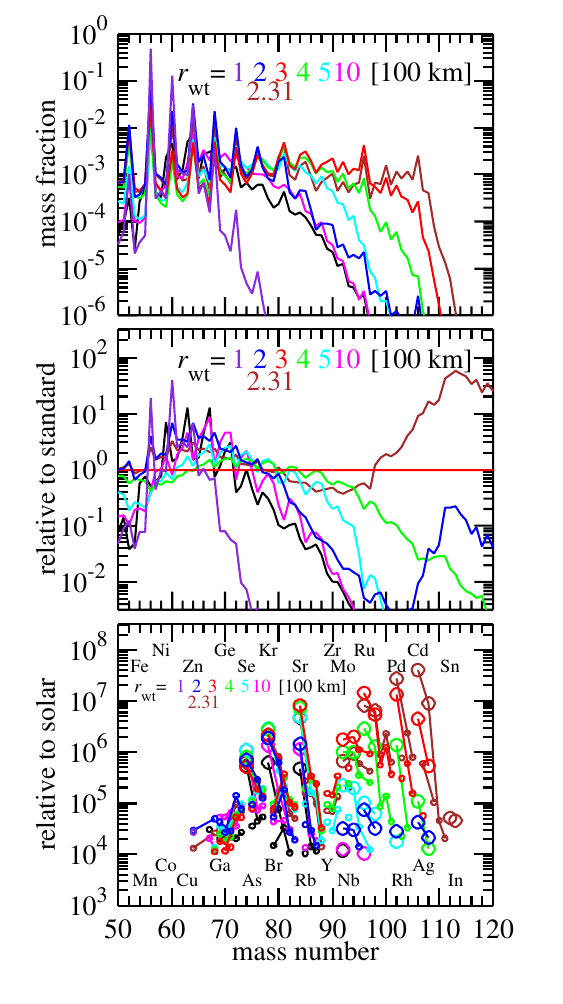}
    \includegraphics[width=0.46\textwidth]{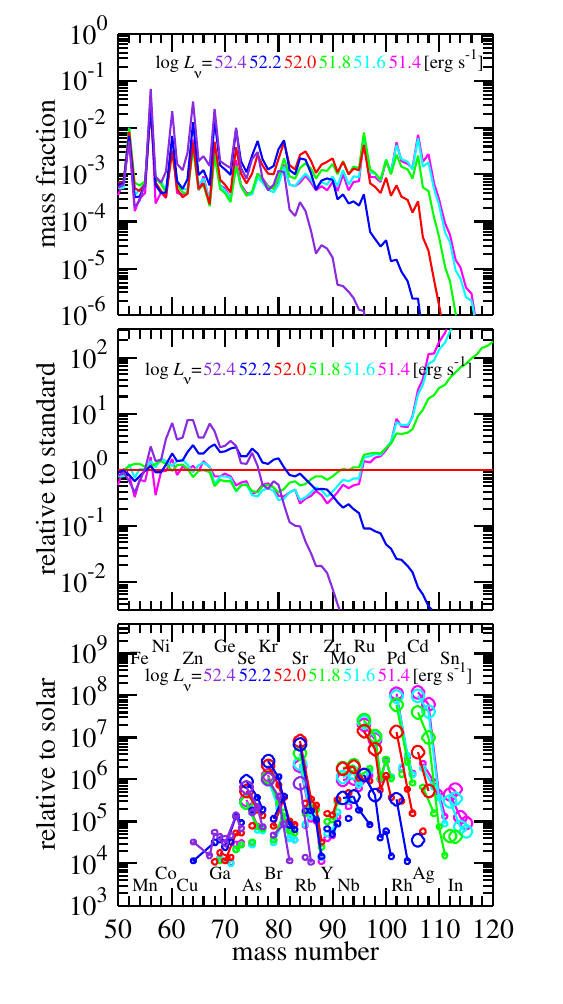}
    \caption{Comparison of nucleosynthesis results with $Y_\mathrm{e,3} = 0.55$ for various wind-termination radii $r_\mathrm{wt}$ (left) and neutrino luminosities $L_\mathrm{\nu}$ (right). The fixed values of $L_\nu = 10^{52}$~erg/s and $r_\mathrm{wt} = 300$~km are adopted for the former (left) and the latter (right), respectively. Mass fractions (top) and their ratios (middle) with respect to the reference cases (red) of $r_\mathrm{wt} = 300$~km (left) or $L_\nu = 10^{52}$~erg/s (right) are shown as a function of atomic mass number. The bottom panels show the isotopes (connected by a line for a given element) with respect to their solar values. In the left panel, the result without wind termination is presented in black. The names of elements are indicated in the upper (even $Z$) and lower (odd $Z$) sides at their lowest mass numbers. (\citealt{Wanajo2011b}; reproduced with permission of the AAS.)}
    \label{fig:abun1}
\end{figure}

The top-left panel of Fig.~\ref{fig:abun1} shows the mass fractions of synthesized nuclei as a function of atomic mass number with the initial $Y_\mathrm{e}$ that becomes 0.55 at $T_\mathrm{GK} = 3$ (the beginning of a $\nu$p-process, hereafter denoted $Y_\mathrm{e,3}$). As mentioned in the previous section, a $\nu$p-process begins from a seed nucleus of $^{56}$Ni made in NSE and QSE. Neutrino irradiation from a proto-neutron star induces electron anti-neutrino capture on abundant free protons, which releases free neutrons as found in Eq.~(\ref{eq:anucap}). Subsequent (slower) $\beta^+$-decay is replaced by much faster neutron capture in Eq.~(\ref{eq:np}). As a result, these successive proton and neutron captures bring the seed nuclei to higher mass numbers through the unstable proton-rich region. However, as can be seen in Fig.~\ref{fig:abun1} (top-left), the heaviest species produced by $\nu$p-processing is highly dependent on the location of wind termination $r_\mathrm{rt}$. Compared to the case without wind termination (black), the $\nu$p-process becomes more efficient with decreasing $r_\mathrm{rt}$ down to 231~km and then less efficient for $r_\mathrm{rt} < 231$~km (see also the middle-left panel in Fig.~\ref{fig:abun1}). This is a consequence of the fact that the duration of the $\nu$p-process (yellow band in Fig.~\ref{fig:abun1}, bottom left) becomes longer with the presence of wind termination. However, wind termination at smaller $r_\mathrm{wt}$ also leads to a longer duration of the seed ($^{56}$Ni) production in QSE and by an $\alpha$-process, resulting in a less efficient $\nu$p-process owing to a fewer number of free protons per $^{56}$Ni. 

The right panels in Fig.~\ref{fig:abun1} compare the results with various neutrino luminosities $L_\nu$ but a fixed $r_\mathrm{rt} = 300$~km. It is found that the $\nu$p-process becomes more efficient for a lower $L_\nu$ and thus at late times. This is principally due to a higher entropy for a lower $L_\nu$ (or at late times; see the middle-left panel in Fig.~\ref{fig:property}), which leads to a higher proton-to-seed ratio at the onset of a $\nu$p-process. However, the early neutrino-heated ejecta should have moved away from the proto-neutron star to a large distance at late times and thus wind termination takes place at a sizably larger $r_\mathrm{wt}$. These counterbalancing effects of $r_\mathrm{wt}$ and $L_\nu$ suggest a maximal efficiency of a $\nu$p-process with a lower $L_\nu$ but a sufficiently smaller $r_\mathrm{wt}$ ($<$ a few 100~km). 

\begin{figure}
    \centering
    \includegraphics[width=0.485\textwidth]{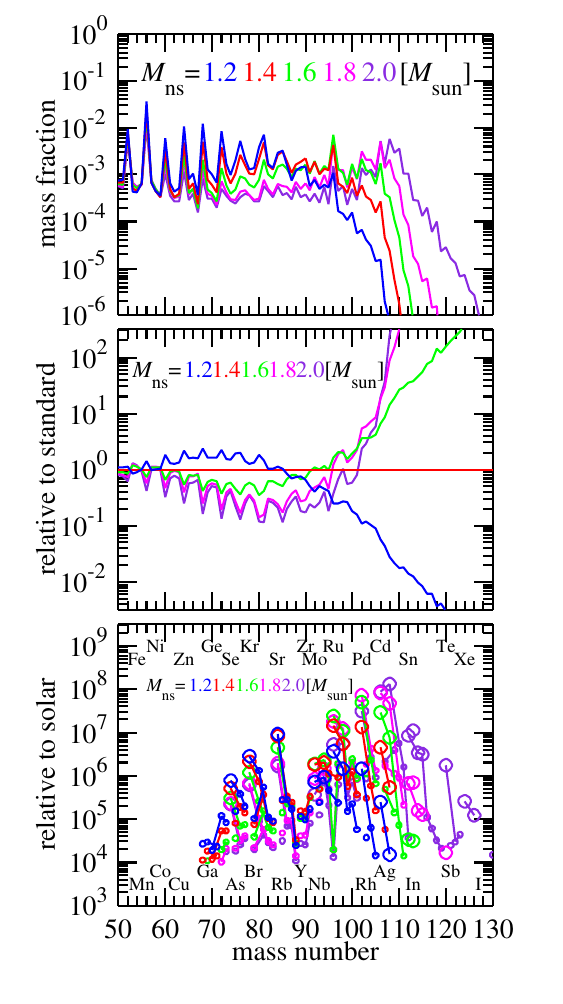}
    \includegraphics[width=0.46\textwidth]{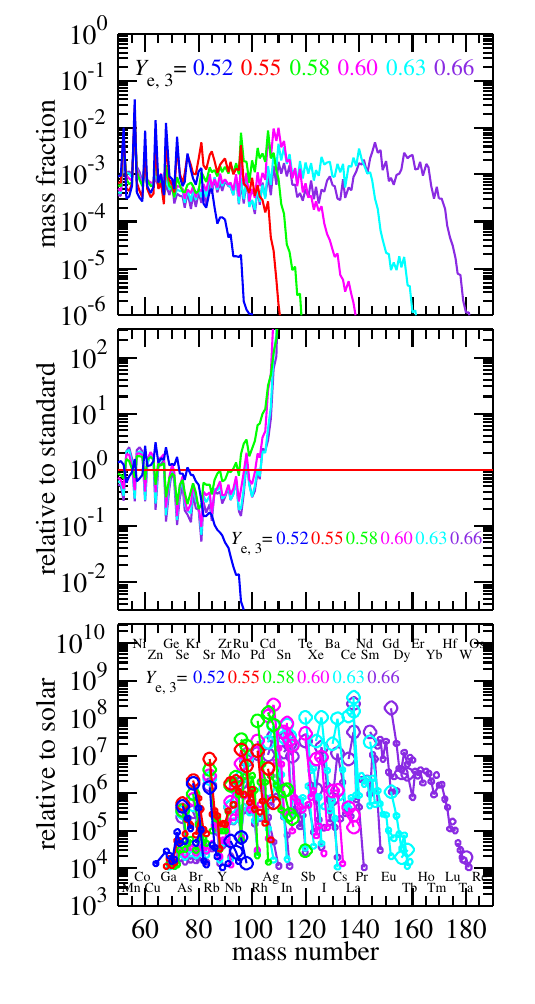}
    \caption{Same as Fig.~\ref{fig:abun1}, but for various neutron star masses (left) and electron fractions. (\citealt{Wanajo2011b}; reproduced with permission of the AAS.)}
    \label{fig:abun2}
\end{figure}

Fig.~\ref{fig:abun2} displays the results with various neutron star masses (left) and $Y_\mathrm{e,3}$ (right) but with fixed $Y_\mathrm{e,3}$ $= 0.55$ (left) and $M = 1.4\, M_\odot$ (right). For both panels, the neutrino luminosity and the wind-termination radius are set to be $L_\nu = 10^{52}$~erg/s and $r_\mathrm{rt} = 300$~km, respectively. Similar to the neutron-rich case in the previous subsection, the neutrino-driven winds from a more massive neutron star result in the production of heavier nuclei as can be seen in the left panel of Fig.~\ref{fig:abun2}. This is due to the higher entropy and the shorter expansion timescale for a greater $M$ (middle panels in Fig.~\ref{fig:property}), which leads to a higher proton-to-seed ratio at the onset of a $\nu$p-process. More proton-rich winds (i.e., with a higher $Y_\mathrm{e}$) also result in the production of heavier isotopes (Fig.~\ref{fig:property}, right). This is a consequence of the fact that the proton-rich matter consists predominantly of free protons, $\alpha$ particles, and $^{56}$Ni at the time at which freezeout from QSE occurs (see Figs.~\ref{fig:freeze} and \ref{fig:toseed}), where the proton-richness is almost entirely due to the presence of free protons. As a result, the proton-to-seed ratio at the beginning of a $\nu$p-process becomes higher for a greater $Y_\mathrm{e}$.

The bottom panels in Figs.~\ref{fig:abun1} and \ref{fig:abun2} show the abundance ratios of synthesized isotopes with respect to their solar values \citep{Lodders2009}, or production factors (with quite huge values due to the omission of outer envelopes). As can be seen, proton-rich isotopes for given elements such as $^{74}$Se, $^{78}$Kr, $^{84}$Sr, $^{92,94}$Mo, $^{96,98}$Ru, $^{102}$Pd, and $^{106,108}$Cd, known as p-nuclei, exhibit large production factors. This is due to the fact that the $\nu$p-process proceeds through the unstable proton-rich region. After the end of $\nu$p-processing, the synthesized nuclei decay back to the proton-rich side of $\beta$-stability. However, the production of proton-rich isotopes is generally limited up to $A \sim 110$ (except for very proton-rich cases with $Y_\mathrm{e,3} > 0.6$) because of higher Coulomb barriers for larger atomic numbers. Therefore, heavier p-nuclei likely originate from other sources such as $\gamma$-process \citep{Rayet1995,Rauscher2002} and $\nu$-process \citep{Woosley1990} in core-collapse supernovae. Note that light p-nuclei up to $^{92}$Mo can also be produced in the early neutrino-heated ejecta of core-collapse supernova (see the results for $15\, M_\odot$ and $27\, M_\odot$ stars in Fig.~\ref{fig:pfiso}). Thus, the innermost ejecta of core-collapse supernova, that is, neutrino-heated ejecta and neutrino-driven winds, are likely sites of p-nuclei with $A = 74$--108.

\section{\textit{Summary}}
\label{sec:summary}

The innermost ejecta of core-collapse supernovae are likely sites of some heavy nuclei. Recent hydrodynamical studies suggest that the early neutrino-heated ejecta are modestly neutron-rich or proton-rich ($Y_\mathrm{e} \sim 0.4$--0.6). Under such conditions, a variety of light trans-iron isotopes with $Z \sim 30$--40 (from Zn to Zr) are synthesized predominantly in nuclear equilibrium (NSE and QSE), although the amounts produced and their distributions are highly dependent on the progenitor masses. In addition to these isotopes, the neutrino-heated ejecta for low-mass and massive progenitors can be possible sources of a neutron-rich isotope $^{48}$Ca and a proton-rich isotope $^{92}$Mo, respectively, for which their origins are currently unknown. Subsequent neutrino-driven winds are unlikely the sources of heavy r-process nuclei because of their expected modest neutron-richness ($Y_\mathrm{e} > 0.4$), although a weak r-process may synthesize light r-process nuclei up to $A \sim 110$. In light of recent hydrodynamical studies showing proton-richness, neutrino-driven winds are likely sites of the $\nu$p-process that synthesizes proton-rich nuclei up to $A \sim 110$. In summary, together with the early neutrino-heated matter and the subsequent neutrino-driven winds, the innermost ejecta of core-collapse supernovae can be important sources of isotopes with $Z \sim 30$--40 as well as p-nuclei $^{74}$Se, $^{78}$Kr, $^{84}$Sr, $^{92,94}$Mo, $^{96,98}$Ru, $^{102}$Pd, and $^{106,108}$Cd. The overall picture of such nucleosynthesis appears quite robust at qualitative levels. However, more progress in astrophysical modeling as well as in relevant nuclear physics (both experimental and theoretical) will be needed for more quantitative predictions of nucleosynthetic yields from the innermost regions of core-collapse supernovae.


\newcommand{\aap}{Astronomy \& Astrophysics}
\newcommand{\apj}{Astrophysical Journal}
\newcommand{\apjl}{Astrophysical Journal Letters}
\newcommand{\apjs}{Astrophysical Journal Supplements}
\newcommand{\apss}{Astrophysics and Space Science}
\newcommand{\mnras}{Monthly Notices of the Royal Astronomical Society}
\newcommand{\prd}{Physical Review D}
\newcommand{\prl}{Physical Review Letters}
\newcommand{\nat}{Nature}

\bibliography{reference}

\end{document}